\documentclass{jfm}
\usepackage{graphicx,color}
\usepackage{epstopdf, epsfig}

\definecolor{cinnamon}{rgb}{0.82, 0.41, 0.12}
\def\MN#1{{\textcolor{black}{#1}}}         

\shorttitle{Turbulence modulation by finite-size spheroidal particles}
\shortauthor{M. N. Ardekani and L. Brandt}

\title{Turbulence modulation in channel flow of finite-size spheroidal particles}

 \author{M. Niazi Ardekani\aff{1} \corresp{\email{mehd@mech.kth.se}}
 \and L. Brandt\aff{1}}

\affiliation{\aff{1} Linn\'e Flow Centre and SeRC (Swedish e-Science Research Centre), KTH Mechanics, \\ SE-100 44 Stockholm, Sweden}

\begin{document}

\maketitle

\begin{abstract}

Finite-size particles modulate wall-bounded turbulence, leading for the case of spherical particles to increased drag also owing to the formation of a particle wall layer. Here, we study the effect of particle shape on the turbulence in suspensions of spheroidal particles at volume fraction $\phi = 10\%$ and show how the near-wall particle dynamics deeply changes with the particle aspect ratio and how this affects the global suspension behavior.
Direct numerical simulations are performed using a direct-forcing immersed boundary method to account for the dispersed phase, combined with a soft-sphere collision model and lubrication corrections for short-range particle-particle and particle-wall interactions. 
\\ 
The turbulence reduces with the aspect ratio of oblate particles, leading to drag reduction with respect to the single phase flow for particles with aspect ratio $\mathcal{AR}\leq1/3$, \MN{when the significant reduction in Reynolds shear stress is more than the compensation by the additional stresses, induced by the solid phase.
Oblate particles are found to avoid the region close to the wall, travelling parallel to it with small angular velocities, while preferentially sampling high-speed fluid in the wall region. 
Prolate particles, also tend to orient parallel to the wall and avoid its vicinity. Their reluctancy to rotate around spanwise axis reduce the wall-normal velocity fluctuation of the flow and therefore the turbulence Reynolds stress similar to oblates; however, they undergo rotations in wall-parallel planes which increases the additional solid stresses due to their relatively larger angular velocities compared to the oblates. These larger additional stresses compensates for the reduction in turbulence activity and leads to a wall-drag similar to that of single-phase flows. Spheres on the other hand, form a layer close to the wall with large angular velocities in spanwise direction, which increases the turbulence activity in addition to exerting the largest solid stresses on the suspension, in comparison to the other studied shapes. Spherical particles therefore increase the wall-drag with respect to the single-phase flow.}             
\end{abstract}
\section{Introduction}
The presence of solid rigid particles in a suspension alters the global transport and rheological properties of the mixture in complex, and often unpredictable, ways. These suspensions are relevant as found in many environmental and industrial processes such as sediment transport in estuaries \citep{Mehta2014}, blood flow in the human body, pyroclastic flows and pulp fibers in paper making industry \citep{Lundell2011}. 

When the Reynolds number is sufficiently high, the flow becomes turbulent, with chaotic and multi-scale dynamics. In this regime, any solid object larger than the smallest scales of the flow can alter the turbulent structures at or below its size  \citep{Naso2010}, leading to turbulence modulation at large enough volume fractions \citep{Lucci2010,Tanaka2015}. 
Particulate wall-bounded turbulent flows in the inertial regime have been investigated in recent years mainly for spherical particles \citep{Matas2003,Loisel2013,Yu2016,Lashgari2014,Wang2016}. \cite{Picano2015} investigated dense suspensions in turbulent channel flow up to volume fraction of $20\%$. Their study revealed that the overall drag increase is due to the enhancement of the turbulence activity up to a certain volume fraction and to the particle-induced stresses at higher concentrations. \cite{Costa2016} explained that the turbulent drag of sphere suspensions is always higher than what predicted by only accounting for the effective suspension viscosity. They attributed this increase to the formation of a particle wall layer, a layer of spheres forming near the wall in turbulent suspensions. Based on the thickness of the particle wall layer, they proposed a relation able to predict the friction Reynolds number as function of the bulk Reynolds number.     

The dynamics of suspension in the presence of finite-size rigid non-spherical particles are less understood \citep{Prosperetti2015}, with just a few studies in recent years \citep{Do2014,Mehdi2017,Eshghinejadfard2017}. In our previous study \citep{Mehdi2017}, we performed simulations of turbulent channel flow with oblate spheroids of aspect ratio (polar over equatorial radius), $\mathcal{AR} = 1/3$, up to volume fraction $\phi = 15 \%$.
We demonstrated that oblate particles induce an overall drag reduction as the volume fraction $\phi$ increases within the investigated range. Turbulence attenuation was also observed, as opposed to suspensions of spherical particles whose presence increases the turbulence activity for volume fractions below $\phi = 20 \%$. Two main mechanisms are found responsible for drag reduction: i) the absence of a particle wall layer, found to be responsible for the increased dissipation in suspensions of spheres; ii) the tendency of oblate particles to remain with their major axes parallel to the wall in its vicinity, thus shielding the near-wall region from the bulk flow. 

\MN{\cite{Wang20172} studied spheroidal particles in a turbulent plane Couette flow with aspect ratios ranging form $0.5$ to $2$ at $5\%$ volume fraction, revealing that the particle distribution of spheroids in the flow is not significantly modified by their shapes. \cite{Eshghinejadfard2017} performed simulations of prolate particles up to $1.5\%$ volume fraction in a turbulent channel flow, showing the disappearance of the particle wall-layer, previously observed for spheres \citep{Picano2015} and also slight turbulence attenuation with respect to the single-phase flow. Finite-size rigid fibers are studied in the work of \citep{Do2014} at low volume fractions around $0.1\%$. These authors report particle sampling of high speed regions close to the wall and also turbulence attenuation with increasing fiber length.} 

Given the importance of particle dynamics on near-wall turbulence, we further investigate the shape effect by simulating spheroids of different aspect ratios $\mathcal{AR}$  at $\phi = 10 \%$, comparing prolates ($\mathcal{AR} > 1$), spheres ($\mathcal{AR} = 1$) and oblates ($\mathcal{AR} < 1$ ).  

\MN{\section{Methodology}
\subsection{Governing equations}}
In this work we employ the numerical model in \cite{Ardekani2016,Mehdi2017} to study dense suspensions of spheroidal particles. 
This approach is based on the immersed boundary method (IBM) initially proposed by \cite{Uhlmann2005,Breugem2012} for spherical particles, extended to spheroids as described in \cite{Ardekani2016}, including lubrication, friction and collision models for the short-range particle interactions. In this scheme, the flow field is resolved on a uniform ($ \Delta x = \Delta y = \Delta z$), staggered, Cartesian grid while particles are represented by a set of Lagrangian points, uniformly distributed on the surface of each particle. 
\\
\MN{The incompressible Navier-Stokes equations describe the flow field in the Eulerian phase: 
\begin{eqnarray}
\label{eq:NS1}  
\rho_f (\frac{\partial \textbf{u}}{\partial t} + \textbf{u} \cdot \nabla \textbf{u} ) &=&  -\nabla p - \nabla p_e + \mu_{f} \nabla^2 \textbf{u} + \rho_f \textbf{f} \, , \\ [8pt]
\nabla \cdot \textbf{u} &=& \, 0 \, .
\label{eq:NS2} 
\end{eqnarray}
where $\textbf{u}$ is the fluid velocity, $\nabla p_e$ is the external pressure gradient that drives the flow with a constant bulk velocity ($U_b$) of the combined phase,  $p$ is the modified pressure (the total pressure minus $p_e$) and $\rho_f$ and $ \mu_{f}$ are the density and dynamic viscosity of the fluid. The additional term $\textbf{f}$ on the right hand side of equation (\ref{eq:NS1}) is the IBM force field, active in the immediate vicinity of a particle surface to enforce no-slip and no-penetration boundary conditions. 
\\
The point force $F_l$ is calculated first at each Lagrangian point using the difference between the particle surface velocity and the interpolated first prediction velocity at the same point.  The first prediction velocity is obtained by advancing equation (\ref{eq:NS1}) in time, using an explicit low-storage Runge-Kutta method, without considering the force field $\textbf{f}$. The forces, $\textbf{F}_l$, are then integrated to the force field $\textbf{f}$ using the regularized Dirac delta function $\delta_d$ of \cite{Roma1999} and added to the first prediction velocity followed by the pressure-correction scheme used in \cite{Breugem2012} to project the velocity field in the divergence-free space.
\\
Taking into account the inertia of the fictitious fluid phase, trapped inside the particle volumes, the motion of rigid spheroidal particles are described by Newton-Euler Lagrangian equations:
\begin{eqnarray}
\label{eq:NewtonEulerWim1}  
\rho_p V_p \frac{ \mathrm{d} \textbf{u}_{p}}{\mathrm{d} t} &=&  -\rho_f \sum\limits_{l=1}^{N_L} \textbf{F}_l \Delta V_l + \rho_f  \frac{ \mathrm{d}}{\mathrm{d} t} \left( \int_{V_p} \textbf{u} \mathrm{d} V  \right)  + \left( \rho_p - \rho_f \right)V_p\textbf{g} + \textbf{F}_c , \, \\ [8pt] 
\frac{ \mathrm{d} \left( \textbf{I}_p \, \pmb{\Omega}_{p} \right) }{\mathrm{d} t} &=& -\rho_f \sum\limits_{l=1}^{N_L} \textbf{r}_l \times \textbf{F}_l \Delta V_l + \rho_f  \frac{ \mathrm{d}}{\mathrm{d} t} \left( \int_{V_p} \textbf{r} \times \textbf{u} \mathrm{d} V  \right) + \textbf{T}_c  \, .
\label{eq:NewtonEulerWim2}  
\end{eqnarray}
with $\textbf{U}_p$ and  $\pmb{\Omega}_{p}$ the translational and the angular velocity of the particle. $\rho_p$, $V_p$ and $\textbf{I}_p$ are the particle mass density, volume and moment-of-inertia tensor and $\textbf{g}$ is the gravity vector. The first term on the right-hand-side of equations above describes the IBM force as the summation of all the point forces $F_l$ on the surface of the particle, the second term accounts for the inertia of the fictitious fluid phase, trapped inside the particle and $\textbf{F}_c$ and $\textbf{T}_c$ are the force and the torque, acting on the particles, due to the particle-particle (particle-wall) collisions. 
\\
When the distance between particles (or a particle and a wall) are smaller than one Eulerian grid size, the lubrication force is under-predicted by the IBM. To compensate for this inaccuracy and to avoid computationally expensive grid refinements, a lubrication model based on the asymptotic analytical expression for the normal lubrication force between 
unequal spheres \citep{Jeffrey1982} is used; here we approximate the two spheroidal particles with two spheres with same mass and radius corresponding to the local 
curvature at the points of contact. Using these approximating spheres, a soft-sphere collision model with Coulomb friction takes over the interaction when the particles touch. 
The restitution coefficients used for normal and tangential collisions are $0.97$ and $0.1$, with Coulomb friction coefficient set to $0.15$.
More details about the models and validations can be found in \cite{Ardekani2016,Costa2015}.  
}

\subsection{Flow geometry}
We perform direct numerical simulations of pressure-driven, particulate turbulent channel flow in a computational domain of size $L_x=6h$, $L_y=2h$ and $L_z=3h$ in the streamwise, wall-normal and spanwise directions, where $h$ is half the channel height. The bulk velocity $U_b$ is fixed to guarantee a constant bulk Reynolds number $Re_b \equiv 2hU_b/\nu = 5600$, corresponding to \MN{an average} friction Reynolds number of $Re_\tau \equiv u_{\tau}h/\nu \approx180$ for the single-phase flow \MN{\citep{Kim1987} } with $\nu$, the kinematic viscosity of the fluid phase and $u_{\tau}$, the friction velocity.  Different aspect ratios $\mathcal{AR} = 3$, $1$, $1/2$, $1/3$, $1/4$ and $1/5$ are simulated at volume fraction $\phi = 10\%$ and compared to the single-phase flow at same $Re_b$. We consider non-Brownian neutrally-buoyant rigid particles of different shape and same volume, with equivalent diameter $D_{eq}=h/9$, (i.e.\ the diameter of a sphere of same volume). This corresponds to $5000$ particles inside the computational domain at $\phi = 10\%$.  The volume fraction is kept constant as it was shown before that excluded volume effects have the largest role on the turbulence modulation \citep{Fornari2016}.  

\MN{\begin{table}
  \begin{center}
\def~{\hphantom{0}}
  \begin{tabular}{lccccccc}
       \,\,\,\,\,\,\,\,\,  \,Case & $\mathcal{AR} = 1/5$  & $\mathcal{AR} = 1/4$ & $\mathcal{AR} = 1/3$  & $\mathcal{AR} = 1/2$   & $\mathcal{AR} = 1$ & $\mathcal{AR} = 3$   \\[3pt]
       $ \,\,\,\,\,\,\,\,\,\,\,\, \, \,  \,N_p$  &  \,\,\,\,\,\,\, &  \,\,\,\,\,\,\, &  $5000$ $(\phi = 10\%)$\\[3pt]
      $ \,\,\,\,\,\,\,\,\,\,\,\, \, \,Re_b$  &  \,\,\,\,\,\,\, &  \,\,\,\,\,\,\, &  $5600$\\[3pt]
       $\, L_x \times L_y \times L_z$  &  \,\,\,\,\,\,\, &  \,\,\,\,\,\,\, &  $6h \times 2h \times 3h$\\[3pt]      
       $\,\,\,\,\,\,\,\,\,\,\,\, h / \Delta x$  &  $432$ & $432$ & $288$ & $288$ & $288$ & $432$ \\[3pt]
       $ \,\,\,\,\,\,\,\, {R_L / D_{eq}}$  & $0.855$ & $0.794$ & $0.721$ & $0.630$ & $0.500$ & $1.040$ \\[3pt]          
       $\,\,\,\,\,\,\,\,\,\,\,\, \, \, Re_\tau$  &  $171.5$ & $173$ & $175$ & $182.5$ & $197$ & $179.5$ \\[3pt]       
       $ \,\,\,\,\,\,\,\,\,\,\,\, \,  {\Delta x}^{\,\,+}$  & $0.395$ & $0.401$ & $0.607$ & $0.634$ & $0.684$ & $0.416$ \\[3pt]                   
  \end{tabular}
  \caption{Summary of the simulations, performed in this study. $N_p$ indicates the number of particles with equivalent diameter $D_{eq} = h/9$, corresponding to a $10\%$ volume fraction; $R_L$ is the semi-major axis of spheroidal particles; $Re_\tau$ is the obtained mean friction Reynolds number and ${\Delta x}^{\,\,+}$ is the Eulerian grid size, given in viscous wall units for each simulation.} 
 \label{tab:cases}
 \end{center}
\end{table}}

The simulations are performed with a resolution of $32$ grid points per $D_{eq}$ for the single-phase flow and the cases with $\mathcal{AR} = 1$, $1/2$ \& $1/3$, whereas a resolution of $48$ points per $D_{eq}$ is found necessary to capture the thin boundary layer around the highly curved particles with $\mathcal{AR} = 3$, $1/4$ \& $1/5$. The number of Lagrangian points $N_L$ on the surface of each particle is $N_L = 8122$, $3219$, $3406$, $3720$, $9120$ and $9880$ for the cases with $\mathcal{AR} = 3$, $1$, $1/2$, $1/3$, $1/4$ and $1/5$. \MN{A summary to the simulated cases is given in table~\ref{tab:cases}}.  The simulations start from the laminar Poiseuille flow with random distribution of the particle position and orientation. The noise introduced by the presence of the particles triggers rapidly the transition to a fully turbulent state, after which the statistics are collected for about $16$ large-eddy turnover times $h/u_{\tau}$. \MN{A CFL number of $0.5$ is used for time-stepping to guarantee the numerical stability of the method.}

\section{Results}
In this section we first investigate the effect of particles on the drag by computing the momentum budget, followed by the turbulent kinetic energy (TKE) budget to explain how particles contribute to the production, transfer and dissipation of TKE, key for future modelling. Next, the turbulence modulation is documented by focusing on the fluid phase statistics and, finally, explained in relation to the particle dynamics. 

\subsection{Momentum and TKE budget analysis}
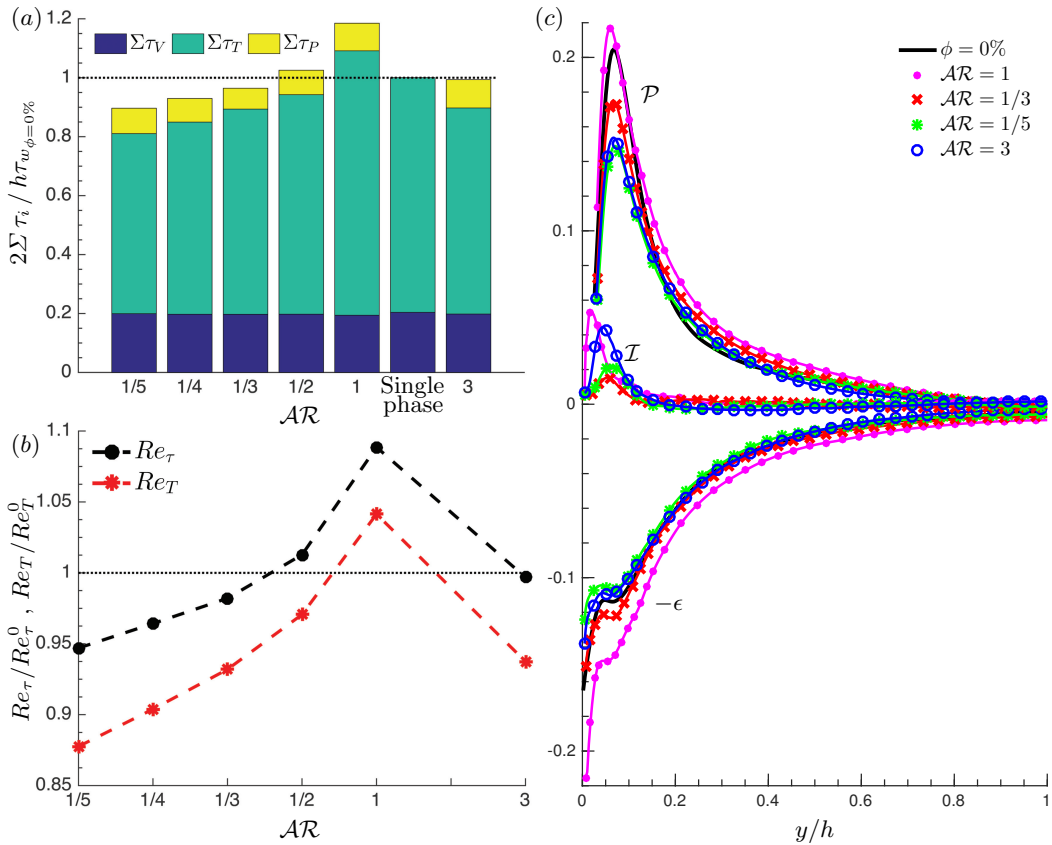
\begin{figure}
  \centering
   \includegraphics[width=0.495\textwidth]{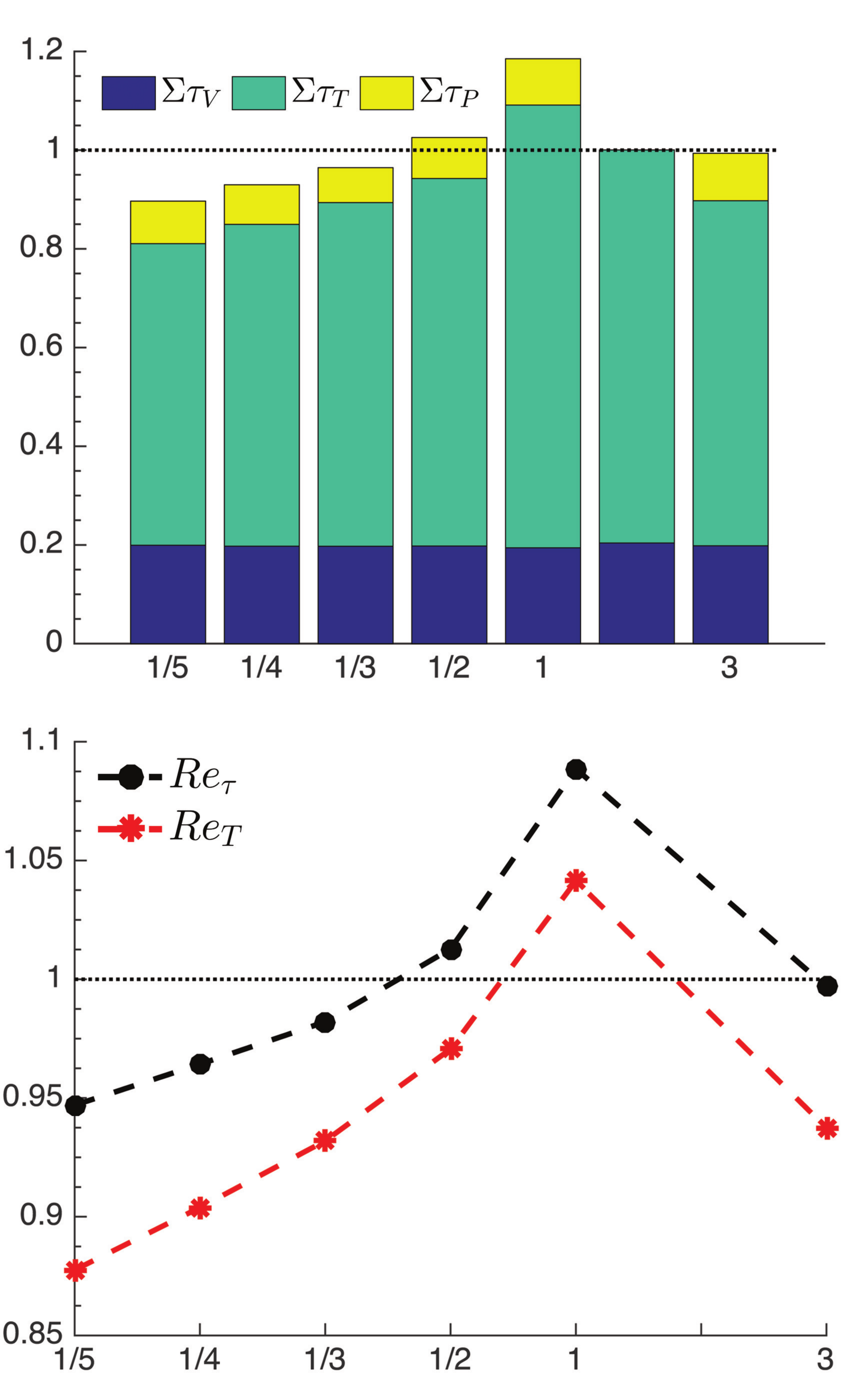}   
   \includegraphics[width=0.495\textwidth]{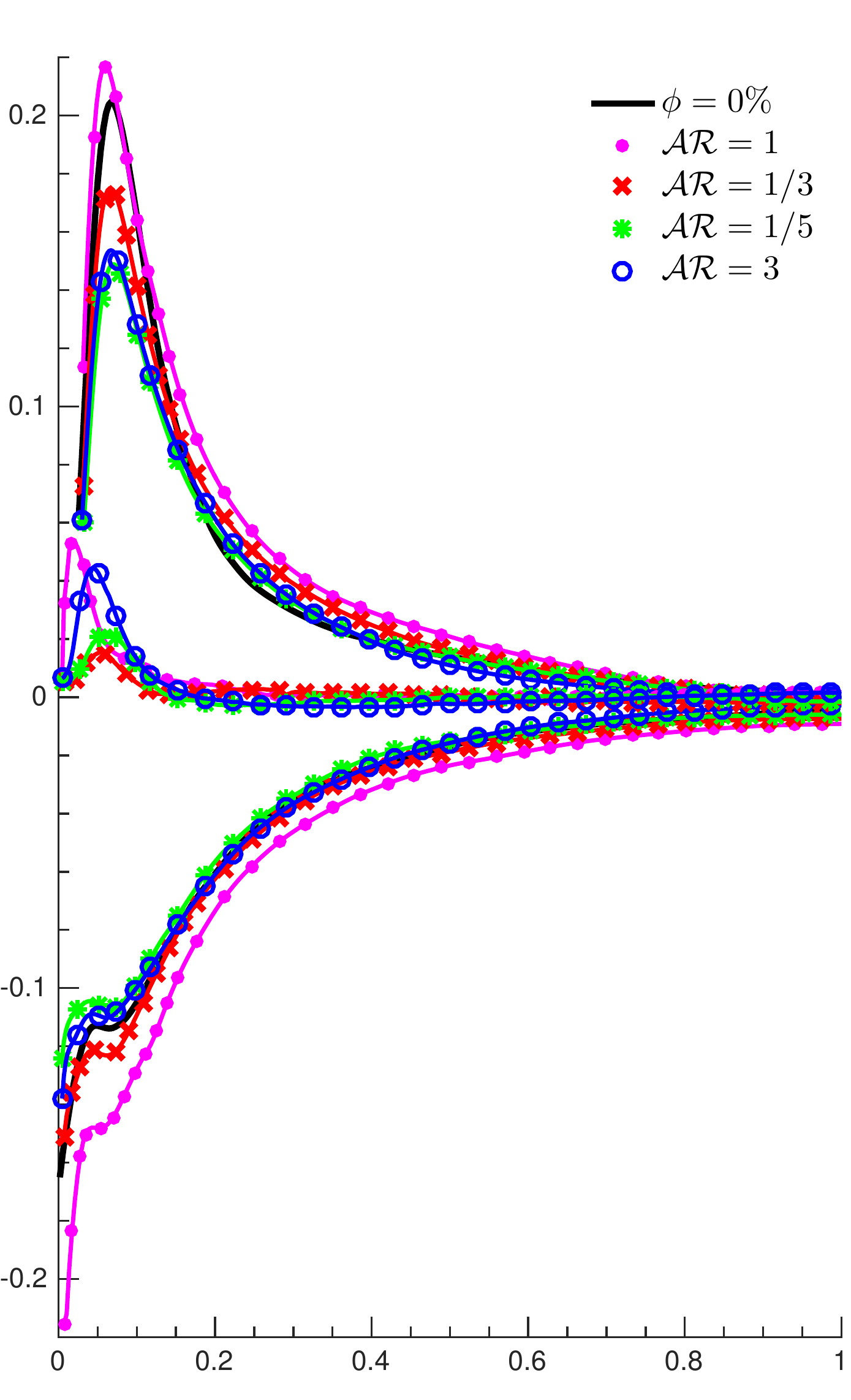}  
   \put(-393,209){\rotatebox{90}{$2 \Sigma \,\tau_{i} \, / \, h \tau_{{w}_{\phi=0\%}}$}}
   \put(-393,38){\rotatebox{90}{$Re_{\tau}/Re^0_{\tau}$~,~$Re_T/Re^0_T$}}       
   \put(-291,150){{$\mathcal{AR}$}}
   \put(-253,161){{Single}}   
   \put(-252,154){{phase}}    
   \put(-291,-5){{$\mathcal{AR}$}}
   \put(-97,-5){{$y / h$}}
   \put(-155,270){{$\mathcal{P}$}}   
   \put(-150,80){{$-\mathcal{\epsilon}$}}     
   \put(-161,173){{$\mathcal{I}$}}     
   \put(-392,299){\footnotesize $(a)$}
   \put(-392,139){\footnotesize $(b)$}
   \put(-195,299){\footnotesize $(c)$}
  \caption{$(a)$: The contribution of the viscous, $\tau_V$,  Reynolds, $\tau_T$, and particle-induced stress, $\tau_P$,  for different aspect ratios, normalized by the drag of the single-phase flow. $(b)$: Turbulence Reynolds number $Re_T$ and friction Reynolds number $Re_{\tau}$, normalized by the values for the single-phase flow case. $(c)$: Production, dissipation and interphase injection of TKE, normalized with ${(u^0_{\tau})}^4/\nu$ of the single phase flow. The production is not depicted in the vicinity of the wall  for clarity.}
\label{fig:budget}
\end{figure}

In particulate turbulent flows, the particle-induced stress also contributes to the mean momentum transfer in the wall-normal direction, in addition to the viscous and Reynolds shear stresses. Based on the formulation proposed by \cite{Zhang2010,Picano2015}, we can write the mean momentum balance in the channel as:
 \begin{equation}
\label{eq:Fl}  
\rho_f u^2_{\tau} \left( 1 - \frac{y}{h} \right ) \, =  \, \mu_f (1-\Phi) \frac{\mathrm{d} U_f}{\mathrm{d} y} \, - \, \rho_f \left[ \, \Phi \langle u^\prime_p v^\prime_p \rangle \, + \, (1-\Phi) \langle u^\prime_f v^\prime_f \rangle \right] \, +  \, \Phi \langle \sigma^p_{xy} \rangle \, \, \, , 
\end{equation}
where the first term on the right hand side is the viscous shear stress, denoted $\tau_V$, the second and the third terms are the turbulent Reynolds shear stress of the combined phase, $\tau_T$, sum of the separate contributions from the two phases, and the fourth term the particle-induced stress $\tau_P$. \MN{$\sigma^p_{xy}$ in the equation above indicates the general stress in the particle phase, normal to streamwise plane and pointing in wall-normal direction, $\Phi$ the mean local solid volume fraction as a function of wall-normal distance $y$, $\rho_f$ the density of the fluid, $U_f$ the mean fluid velocity and $u^\prime$ and $v^\prime$ are the velocity fluctuations in the streamwise and wall-normal directions with the subscripts $f$ and $p$ denoting fluid and particle phases. Note that in the absence of particles ($\Phi=0$), this equation reduces to the classic momentum balance for single-phase turbulent channel flow \citep{Pope2001}. \\
Integrating equation (\ref{eq:Fl}) from $0$ to $h$ and substituting $\rho_f u^2_{\tau}$ with the drag at the wall, $\tau_w$, leads to the contribution of each stress to $\tau_w$, which can be written in discrete form as:
 \begin{equation}
\label{eq:FlInt}  
\tau_w \, =  \, \frac{2}{h} \left[ \Sigma \tau_{V} + \Sigma \tau_{T} + \Sigma \tau_{P} \right] \, \, .
\end{equation}
\\
Each contribution times $2/h$, normalized by the drag of the single-phase flow,  $\tau_{{w}_{\phi=0\%}}$, is given in figure~\ref{fig:budget}$(a)$ for the different particle aspect ratios considered. 
}
An overall drag reduction is observed for the cases with $\mathcal{AR} \leq 1/3$, increasing up to $10\%$ for the smallest $\mathcal{AR}=1/5$.
The largest increase with respect to the unladen case is obtained for spheres, whereas a small drag increase is
still observed for $\mathcal{AR}=1/2$. 
Finally, the drag for the case with prolate particles is almost the same as for the single-phase flow. 
Considering the different contributions, we note the non-monotonic behaviour of the particle-induced stress with decreasing $\mathcal{AR}$ as it reaches a minimum for $\mathcal{AR}=1/3$ before slightly increasing when further decreasing the aspect ratio.
 \MN{The contribution of the viscous shear stress to the total drag is slightly varying among the different aspect ratio cases, while being smaller than the single phase flow due to the presence of $1-\Phi$ in $\tau_{V}$ of particulate cases.}
In addition, we see turbulence attenuation for all non-spherical cases, an effect increasing as the aspect ratio departs from sphericity.
\MN{
The level of turbulence activity is quantified by the turbulence Reynolds number $Re_T\equiv u_T h/\nu$, with $u_T$ the turbulent friction velocity, calculated as the square root of the wall-normal derivative of the Reynolds stress profile for the combined phase, $\langle u^\prime_c v^\prime_c \rangle$, at the centreline of the channel:
 \begin{equation}
\label{eq:UT}  
u_T \, =  \, \sqrt{\mathrm{d} \langle u^\prime_c v^\prime_c \rangle \,/\, \mathrm{d}  \left(y/h\right)} |_{y/h = 1} \, \, .
\end{equation}
}

$Re_T$, normalized by that of the single-phase flow, $Re^0_T \approx 174$, is depicted in figure~\ref{fig:budget}$(b)$ for different $\mathcal{AR}$, together with the ratio $Re_{\tau}/Re^0_{\tau} =Re_{\tau}/180$.  The results indicate that the turbulence activity is decreasing faster than the friction Reynolds number when decreasing the aspect ratio of oblate particles. \MN{This is due to the almost unchanged contribution of viscous shear stress ($ \Sigma \tau_{V}$) and slight increase of particle-induced stress when decreasing the aspect ratio below $\mathcal{AR}=1/3$, meaning that a stronger turbulence attenuation is required in order to further reduce the drag.}

We next consider the turbulent kinetic energy (TKE) budget to quantify the solid-phase contribution to production and dissipation of TKE of the fluid phase.  Considering the Navier-Stokes (NS) equations with the IBM force ($f$), employed to impose no-slip at the particle surface, the TKE budget can be written for a fully developed turbulent channel flow as: 

 \MN{
 \begin{eqnarray}
\label{eq:TKE}  
&-& \frac{\partial \mathcal{T}_j}{\partial x_j} - \mathcal{\epsilon} + \mathcal{P} + \mathcal{I}  = 0 \, \, , \\
&\mathcal{P}& = - \langle u^\prime_i u^\prime_j \rangle \,   \frac{\partial U_i}{\partial x_j} \, \, , \\
&\mathcal{\epsilon} & = \, \nu   \,  \langle \frac{\partial u^\prime_i }{\partial x_j } \,\frac{\partial u^\prime_i }{\partial x_j} \rangle \, \, , \\
&\mathcal{T}_j& = \frac{1}{2} \, \langle u^\prime_i u^\prime_i u^\prime_j \rangle + \frac{1}{\rho} \, \langle u^\prime_j  p^\prime  \rangle - \frac{1}{2} \, \nu \, \frac{\partial \langle u^\prime_i u^\prime_i \rangle}{\partial x_j} \, \, , 
\end{eqnarray}
where the first three terms in equation \ref{eq:TKE} are respectively responsible for the spatial redistribution, viscous dissipation and production of TKE ($0.5\, \langle u^\prime_i u^\prime_i \rangle$) \citep{Mansour1988}, while the fourth term is the interphase energy injection, $\mathcal{I} =  \langle u^\prime_j f_j \rangle$ \citep{Tanaka2015}.  
 \\
 The last three terms in equation \ref{eq:TKE} are calculated for the fluid phase, considering only the points located outside of the volume occupied by the particles (phase-ensemble average), and depicted for different aspect ratios in figure~\ref{fig:budget}$(c)$.} The data indicate a strong reduction of $\mathcal{P}$ in the near-wall region with decreasing $\mathcal{AR}$ for oblate particles. A significant reduction is also observed for prolate particles, while a slight increase with respect to the single-phase flow is seen for spheres. The dissipation term $\mathcal{\epsilon}$ is almost the same as in the single-phase flow  for non-spherical particles, except at the wall where its value reduces with $\mathcal{AR}$. For spherical particles, however, the dissipation is considerably larger throughout the channel. The energy injected by the fluid-solid interactions, i.e. the interphase energy injection at the particle surface, interestingly follows the same trend as the contribution of particle-induced stresses in the momentum balance ($\tau_P $ in figure~\ref{fig:budget}$a$): it reaches a minimum at $\mathcal{AR}=1/3$, before increasing again for lower aspect ratios.  The energy injected by the particles first decreases when decreasing  $\mathcal{AR}$, owing to the reduced rotation rate discussed below; however, it increases again for the smallest aspect ratio considered  ($\mathcal{AR}=1/5$) as the particle surface area increases.
\MN{This can be explained by the fact that a large value of $f$ indicates a force that acts on the fluid and also on the surface of the particle, meaning that in the absence of particle's rigidity, a large deformation of particle surface would have been expected. Indeed the particle's rigidity opposes this deformation and instead increases the particle-induced stress.}

\subsection{Turbulence modulation}
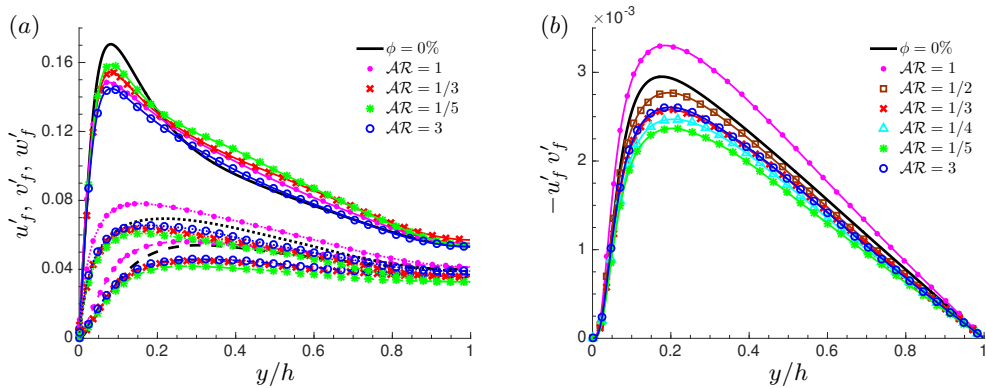
\begin{figure}
  \centering
   \includegraphics[width=0.495\textwidth]{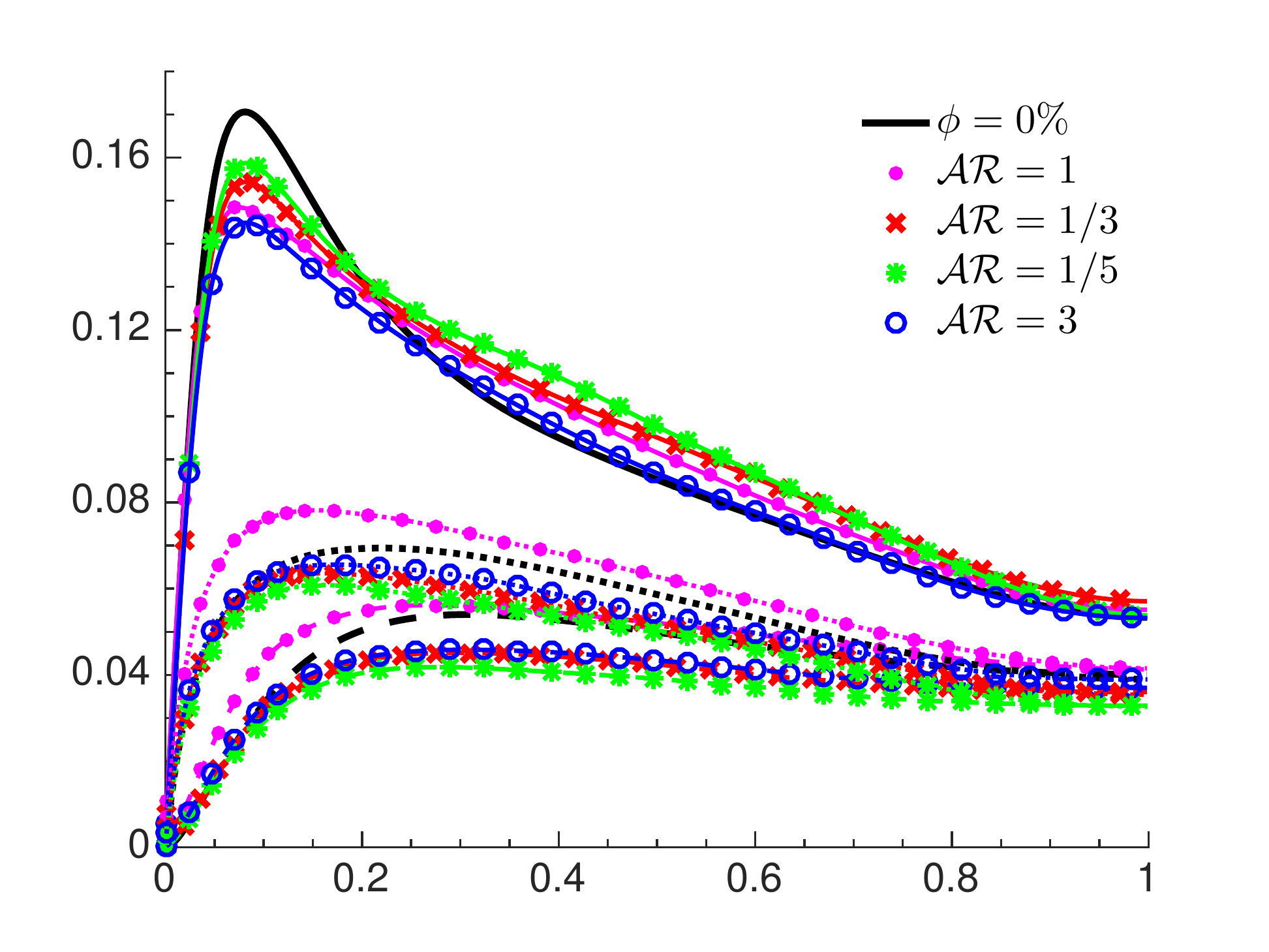}
   \includegraphics[width=0.495\textwidth]{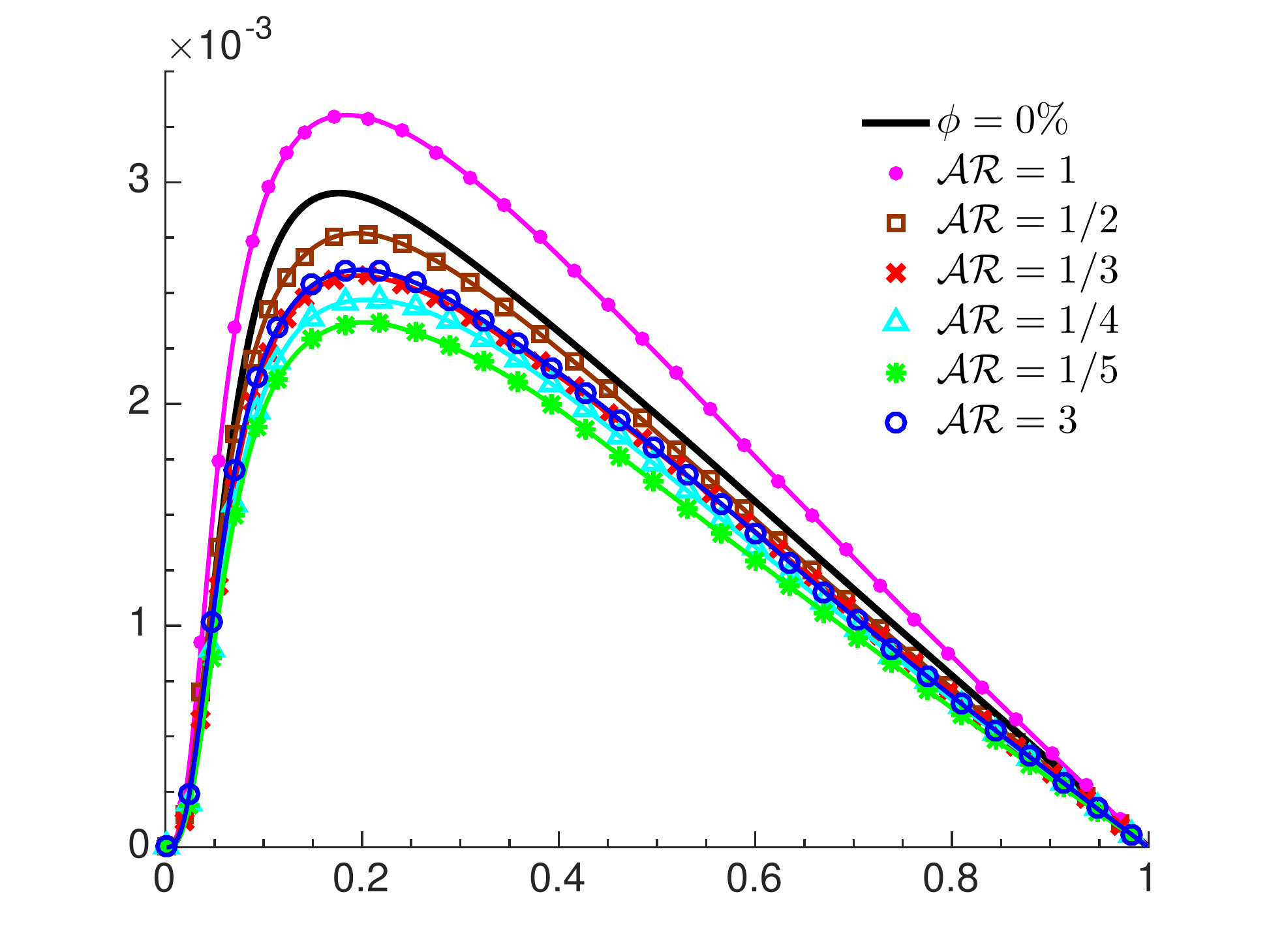}
   \put(-385,53){\rotatebox{90}{$u^\prime_f $, $v^\prime_f $, $w^\prime_f $}}
   \put(-185,62){\rotatebox{90}{$-u^\prime_f \,v^\prime_f$}}
   \put(-99,0){{$y / h$}}
   \put(-292,0){{$y / h$}}
   \put(-385,130){\footnotesize $(a)$}
   \put(-185,130){\footnotesize $(b)$} \\
  \caption{$(a)$: Fluid velocity fluctuations $u^{\prime}$, $v^{\prime}$ and $w^{\prime}$ in the streamwise, ---~, wall-normal, -~-~-~, and spanwise, ...~, directions and $(b)$: Reynolds shear stress, scaled in outer units for different aspect ratios. The velocity fluctuations for the cases with $\mathcal{AR}=1/2$ and $1/4$ are not shown in panel $(a)$ to improve clarity.}
\label{fig:RMS}
\end{figure}

Fluid velocity fluctuations of the fluid phase are depicted in figure~\ref{fig:RMS}$(a)$ for the different particle aspect ratios under investigations. The velocity fluctuations are reduced compared to the single-phase flow for non-spherical particles in the crossflow directions. The effect is observed to increase as the aspect ratio of oblate particles decreases. Spherical particles, on the other hand, increase the crossflow velocity fluctuations. The strongest reduction of $u^{\prime}$ is observed for prolates, followed by spheres despite the slight increase in the core region. Interestingly, $u^{\prime}$ is observed to increase with reducing $\mathcal{AR}$ in the oblate cases. The same behaviour is observed in the literature for other drag reducing flows, e.g.\ polymeric flows, when reduced  spanwise and wall-normal velocity fluctuations are observed \citep{Dubief2004}. Figure~\ref{fig:RMS}$(b)$ shows the Reynolds shear stress of the fluid phase for all studied cases. A significant turbulence attenuation is observed for non-spherical particles, as opposed to the turbulence enhancement for spherical particles. The Reynolds shear stress $-u^\prime_f \,v^\prime_f$ decreases monotonically with decreasing $\mathcal{AR}$ in the oblate cases. Interestingly, the $-u^\prime_f \,v^\prime_f$ pertaining prolates with $\mathcal{AR}=3$ is approximately matching the case of oblates with $\mathcal{AR}=1/3$.   

\begin{figure}
 \centering
   \includegraphics[width=1\textwidth]{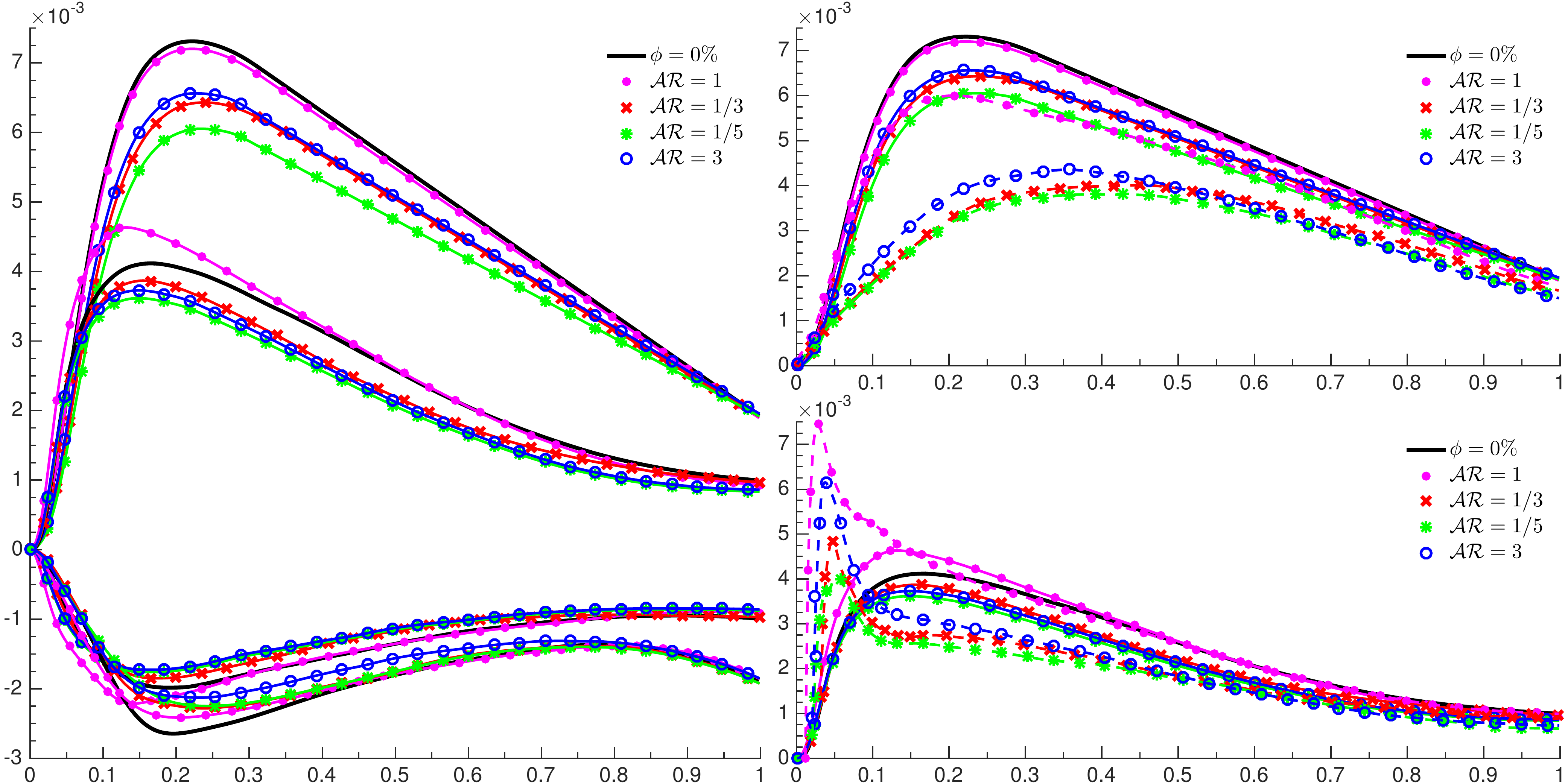}
   \put(-215,103){\Large $Q_2$}  
   \put(-215,80){\Large $Q_4$}   
   \put(-215,48){\Large $Q_1$}   
   \put(-215,16){\Large $Q_3$}    
   \put(-293,-10){{$y / h$}}   
   \put(-102,-10){{$y / h$}}   
   \put(-102,86){{$y / h$}}     
   \put(-55,150){\Large $Q_2$}    
   \put(-55,30){\Large $Q_4$}    
   \put(-393,80){\rotatebox{90}{$-u^\prime_f \,v^\prime_f$} }   
   \put(-390,185){\footnotesize $(a)$}          
   \put(-200,185){\footnotesize $(b)$}  
   \put(-200,92){\footnotesize $(c)$}  
  \caption{$(a)$: Reynolds shear stress, conditionally averaged outside of the particles, for each quadrant of $u^\prime_f \,v^\prime_f$ plane for different aspect ratios.  \MN{$(b)$ - $(c)$: $Q_2$ and $Q_4$ events, conditionally averaged over the fluid inside the spheroidal shells, $5\%$ larger than the particles. The dashed lines in panels $(b)$ and $(c)$, indicate the average over the shells, while the overall averages inside the fluid phase are indicated with the solid lines.}  }
\label{fig:Quad}
\end{figure}
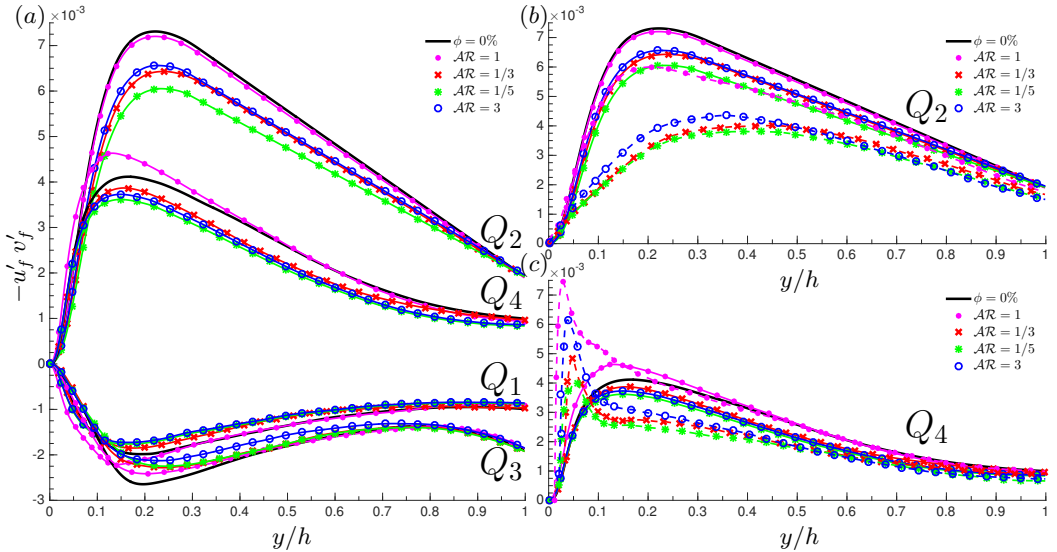

The Reynolds shear stresses are then conditionally averaged over each quadrant of the $u^{\prime}-v^{\prime}$ plane, with $Q_1$ to $Q_4$ referring to the four quadrants.
$Q_2$ (ejections: $u^\prime_f<0$, $v^\prime_f>0$) and $Q_4$ (sweeps: $u^\prime_f>0$, $v^\prime_f<0$) events result in turbulence production, while $Q_1$ and $Q_3$ are responsible for damping. The results of this analysis are depicted in figure~\ref{fig:Quad}$(a)$, where it is observed that the presence of oblate particles reduces the value of $u^{\prime}v^{\prime}$ over all four quadrants, with a far more pronounced dampening of the ejection events. Spherical particles, on the other hand, increase the strength of sweeps close to the wall and reduce the contribution of $Q_3$ events, without altering the others noticeably. Prolate particles exhibit an effect similar to oblates, but with lower dampening of the ejections. 
\MN{Similar analysis is performed, using only the fluid inside the spheroidal shells, $5\%$ larger than the particles. $Q_2$ and $Q_4$ events, conditionally averaged over these shells are depicted in figure~\ref{fig:Quad}$(b)$ and \ref{fig:Quad}$(c)$ (dashed lines), respectively and compared against the obtained results in figure~\ref{fig:Quad}$(a)$. Figure~\ref{fig:Quad}$(b)$ shows that the ejection events are strongly attenuated in the vicinity of the particles compared to the single-phase flow and also compared to the results in figure~\ref{fig:Quad}$(a)$. The attenuation is observed to be stronger for non-spherical particles. Interestingly, the results in figure~\ref{fig:Quad}$(c)$ indicate a region near the wall where the sweeps are strong close to the particles before decreasing below the results of figure~\ref{fig:Quad}$(a)$ in the middle of the channel. The peak is stronger for spherical and prolate particles and it decreases with aspect ratio of oblates. }

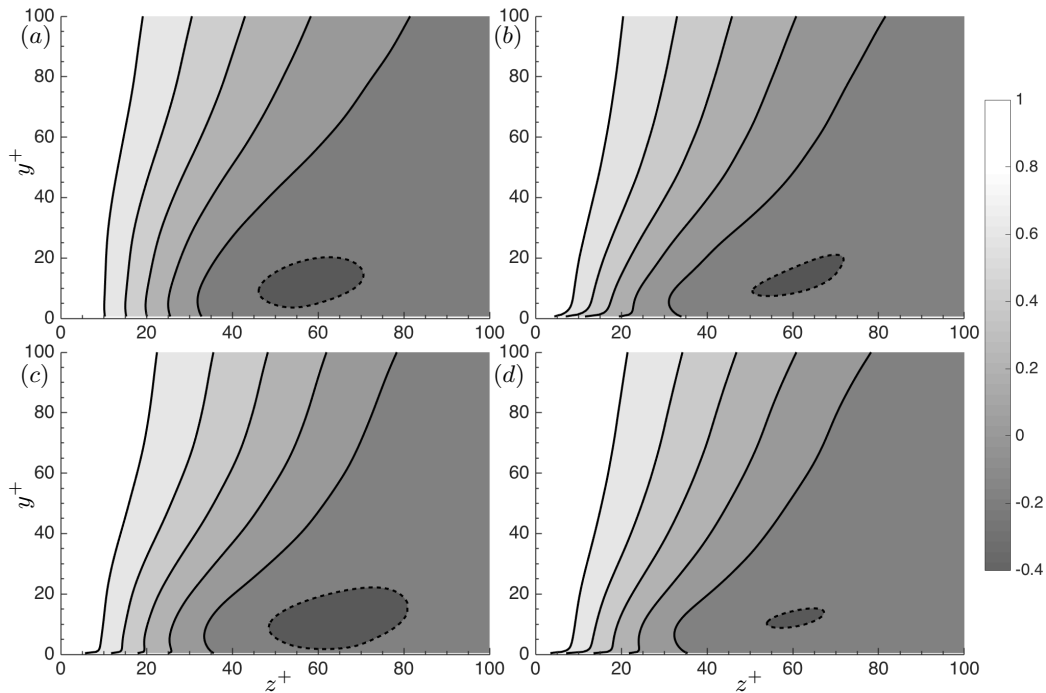
\begin{figure}
 \centering
   \includegraphics[width=1\textwidth]{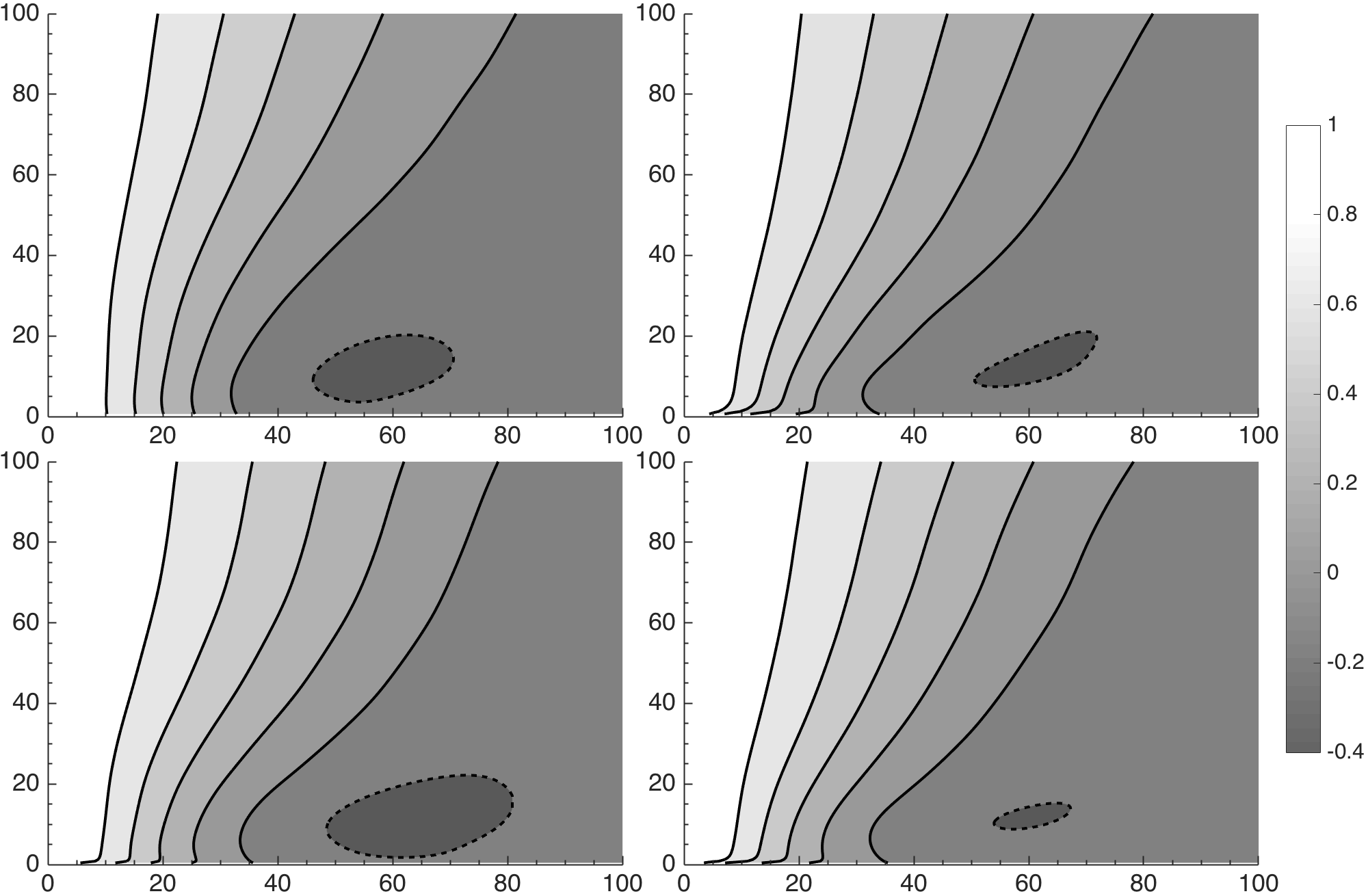}
   \put(-386,238){\footnotesize $(a)$}
   \put(-208,238){\footnotesize $(b)$} 
   \put(-386,111){\footnotesize $(c)$}
   \put(-208,111){\footnotesize $(d)$}       
   \put(-390,185){\rotatebox{90}{ $y^+$} }
   \put(-390,60){\rotatebox{90}{ $y^+$} }
   \put(-297,-5){ $z^+$} 
   \put(-120,-5){ $z^+$}                          
  \caption{Contours of two-point spatial correlation of the streamwise velocity fluctuations as a function of the spanwise spacing for different wall-normal distances: $(a)$ single-phase flow, $(b)$ $\mathcal{AR}=1$, $(c)$ $\mathcal{AR}=1/5$ and $(d)$ $\mathcal{AR}=3$. The inner length scale $\nu/u_{\tau}$, used here, is calculated from the single-phase flow case to allow for a direct comparison. The solid and dashed lines correspond to positive and negative values, ranging from $0.8$ to $-0.2$ with a step of $0.2$.}
\label{fig:Corr}
\end{figure}

\MN{To gain a better understanding of the turbulence structures in the wall region, two-point spatial correlation of the streamwise velocity fluctuations as a function of the spanwise spacing is computed for different wall-normal distances:
\begin{equation}
\label{eq:UT}  
R_{uu} (y,\Delta z) \, =  \, \frac{\langle u^\prime (x,y,z,t) \, u^\prime (x,y,z+\Delta z,t) \rangle}{{u^\prime}^2} \, \, .
\end{equation}
Contours of the one-dimensional autocorrelation of the streamwise velocity fluctuations as a function of the spanwise spacing are given in figure~\ref{fig:Quad} for different aspect ratios. It is well known that the autocorrelations of the streamwise velocity along the spanwise direction in a single-phase turbulent flow  (\ref{fig:Quad}$(a)$), indicate a negative local minimum value in the near wall region around $\Delta z^+ \approx 60$ \citep{Pope2001}. This value indicates half of the spacing between the near-wall streamwise low-speed streaks, characteristic of wall-bounded turbulence. Results in figure~\ref{fig:Quad} indicate a clear increase in the spacing of these structures for the oblate particles with $\mathcal{AR}=1/5$ (\ref{fig:Quad}$(c)$), a characteristic observed in drag-reducing flows \citep{Dubief2004}; however the low-speed streaks are observed to become less correlated for the flow with prolate particles  (\ref{fig:Quad}$(d)$) and to a less extent with spherical ones (\ref{fig:Quad}$(b)$).}

\subsection{Particle dynamics}

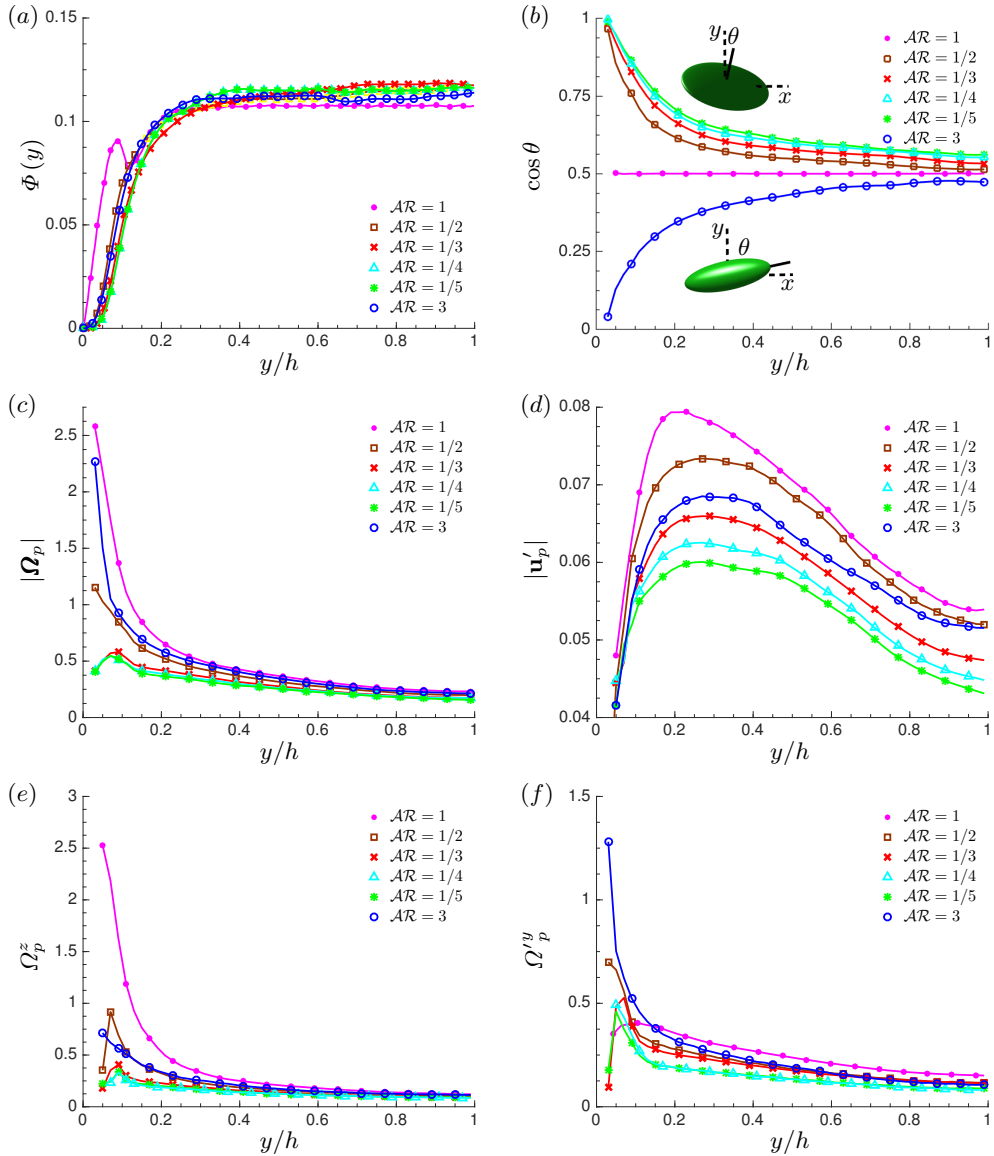
\begin{figure}
  \centering
   \includegraphics[width=0.495\textwidth]{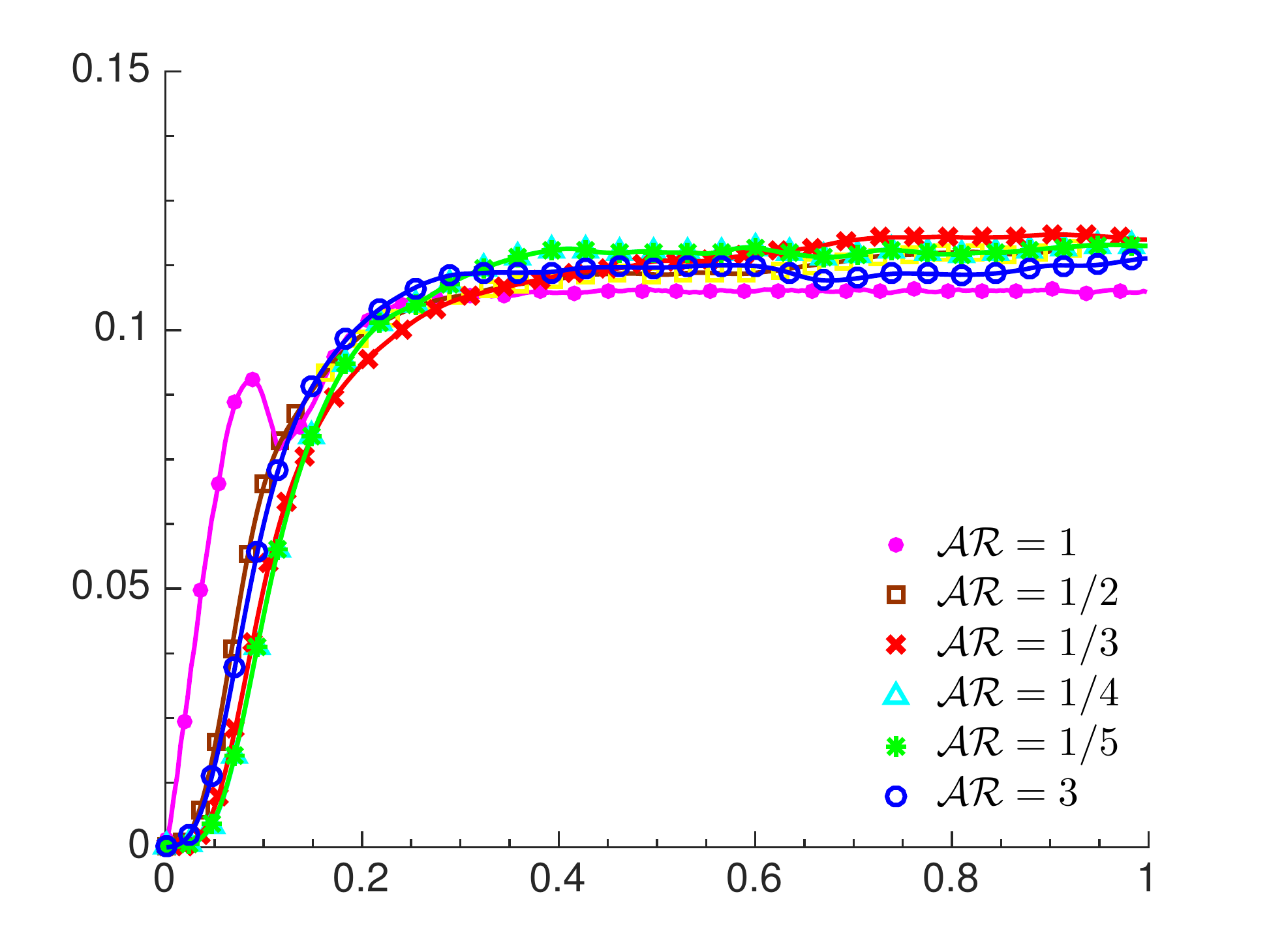}
   \includegraphics[width=0.495\textwidth]{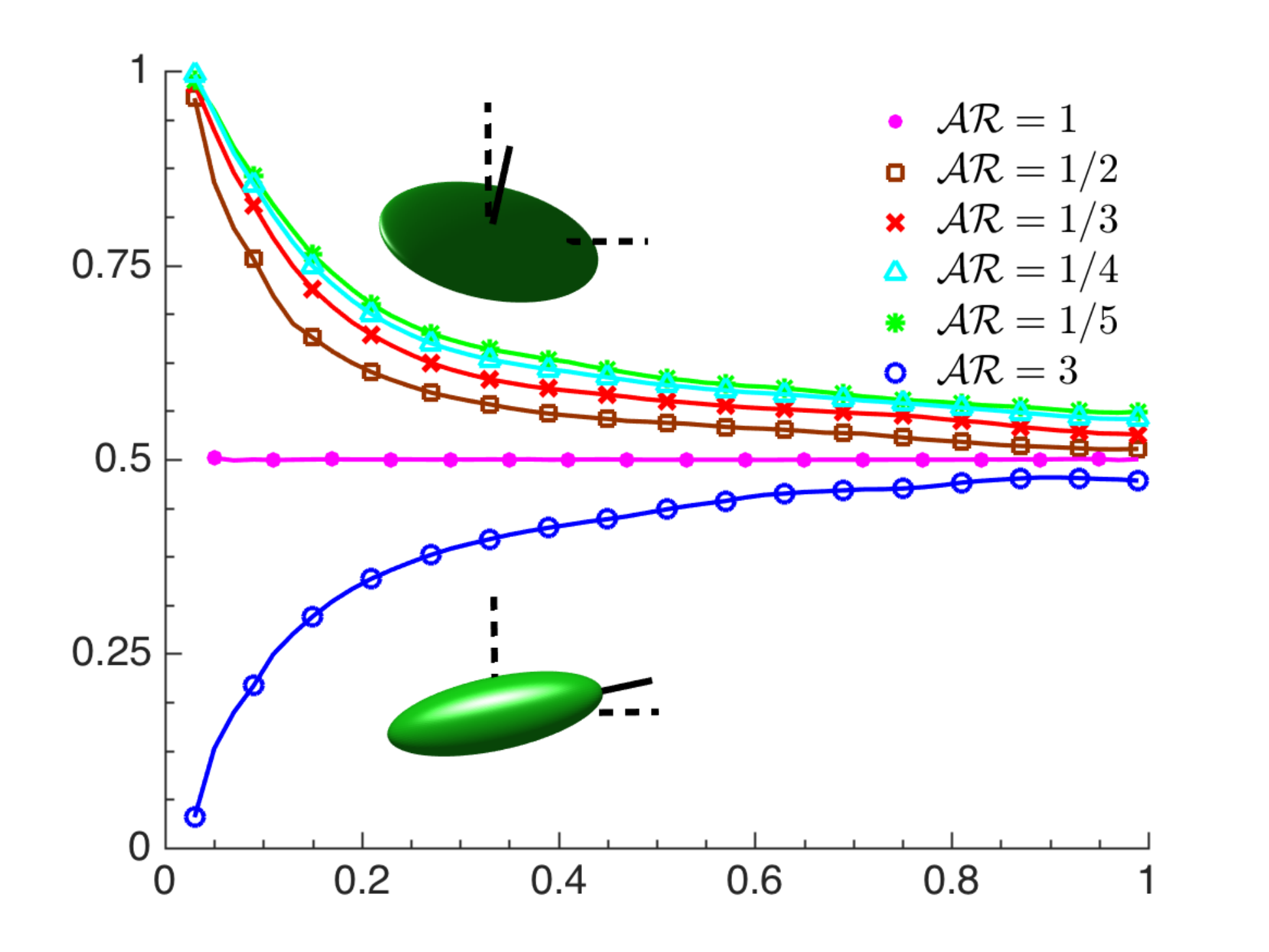}
   \put(-382,67){\rotatebox{90}{$\Phi \, (y)$}}
   \put(-193,67){\rotatebox{90}{$\cos  \theta$}}
   \put(-100,0){{$y / h$}}
   \put(-292,0){{$y / h$}}
   \put(-96,30){{$x$}}
   \put(-122,51){{$y$}}  
   \put(-112,43){{$\theta$}}     
   \put(-97,100){{$x$}}
   \put(-123,125){{$y$}}  
   \put(-116,122){{$\theta$}}    
   \put(-387,130){\footnotesize $(a)$}
   \put(-194,130){\footnotesize $(b)$} \\ 
   \includegraphics[width=0.495\textwidth]{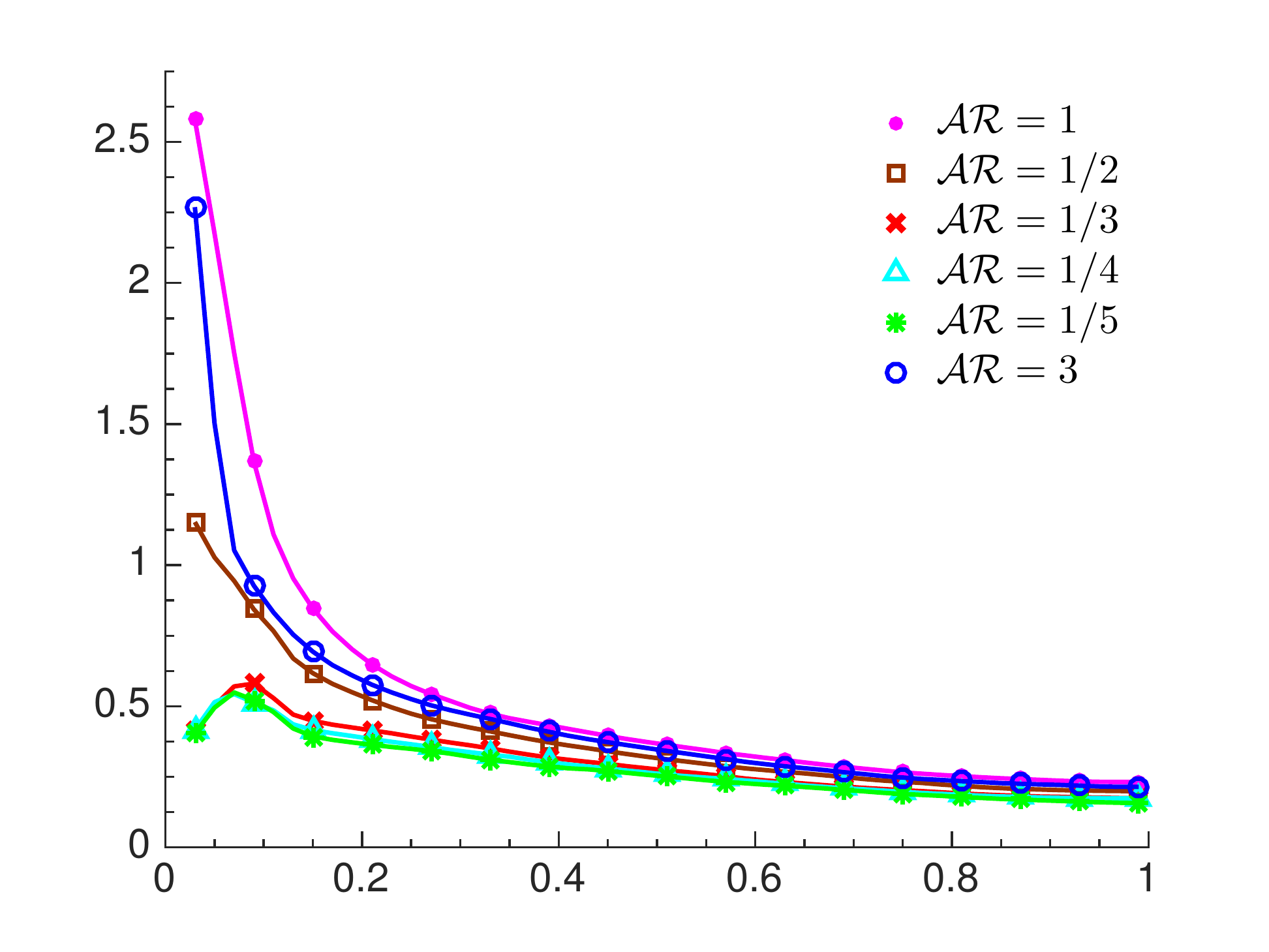}
   \includegraphics[width=0.495\textwidth]{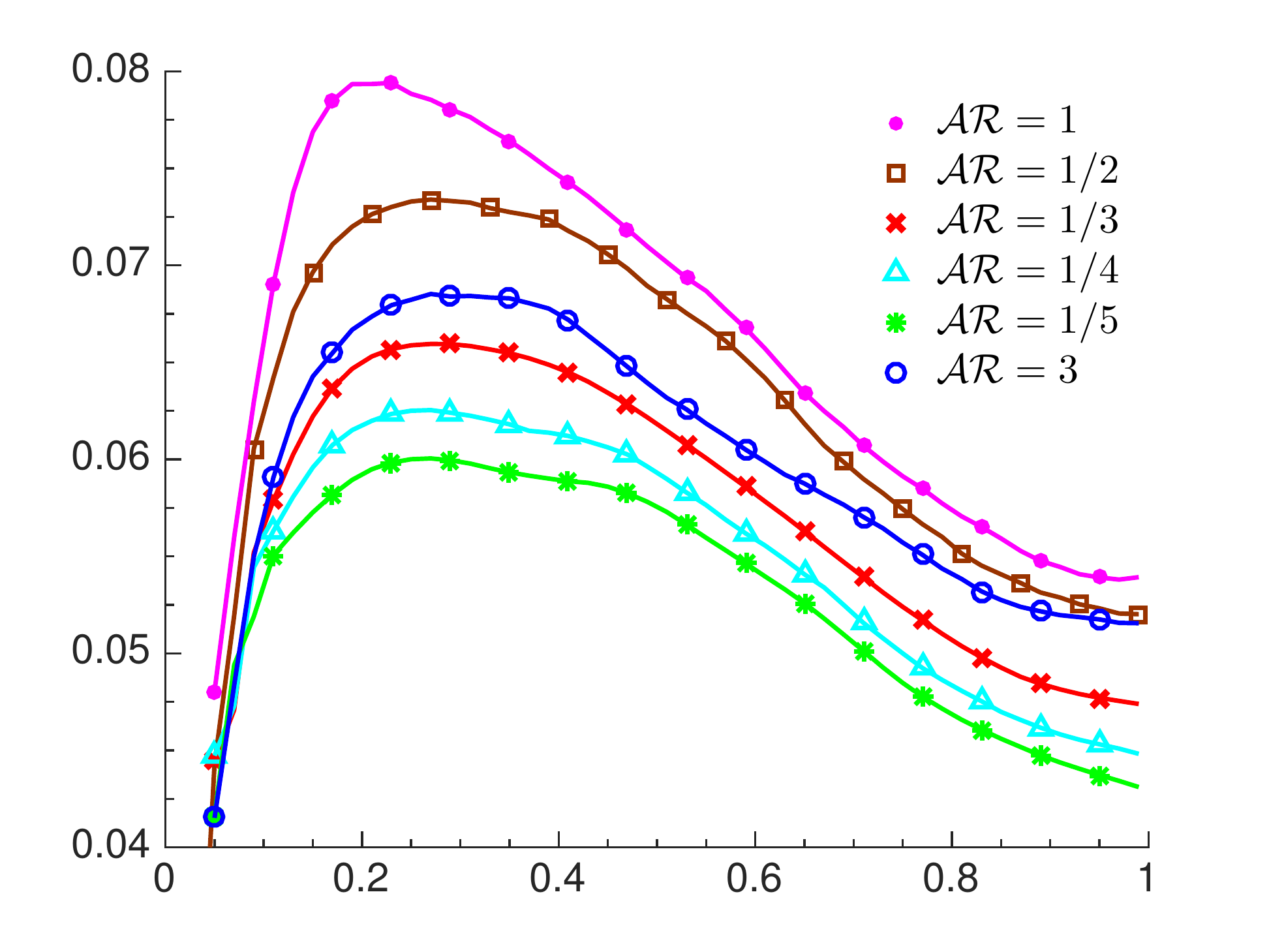}
   \put(-382,66){\rotatebox{90}{$| \pmb{\Omega}_p |$}}
   \put(-193,67){\rotatebox{90}{$| \textbf{u}_p^{\prime} |$}}
   \put(-100,0){{$y / h$}}
   \put(-292,0){{$y / h$}}
   \put(-387,130){\footnotesize $(c)$}
   \put(-194,130){\footnotesize $(d)$} \\
   \includegraphics[width=0.495\textwidth]{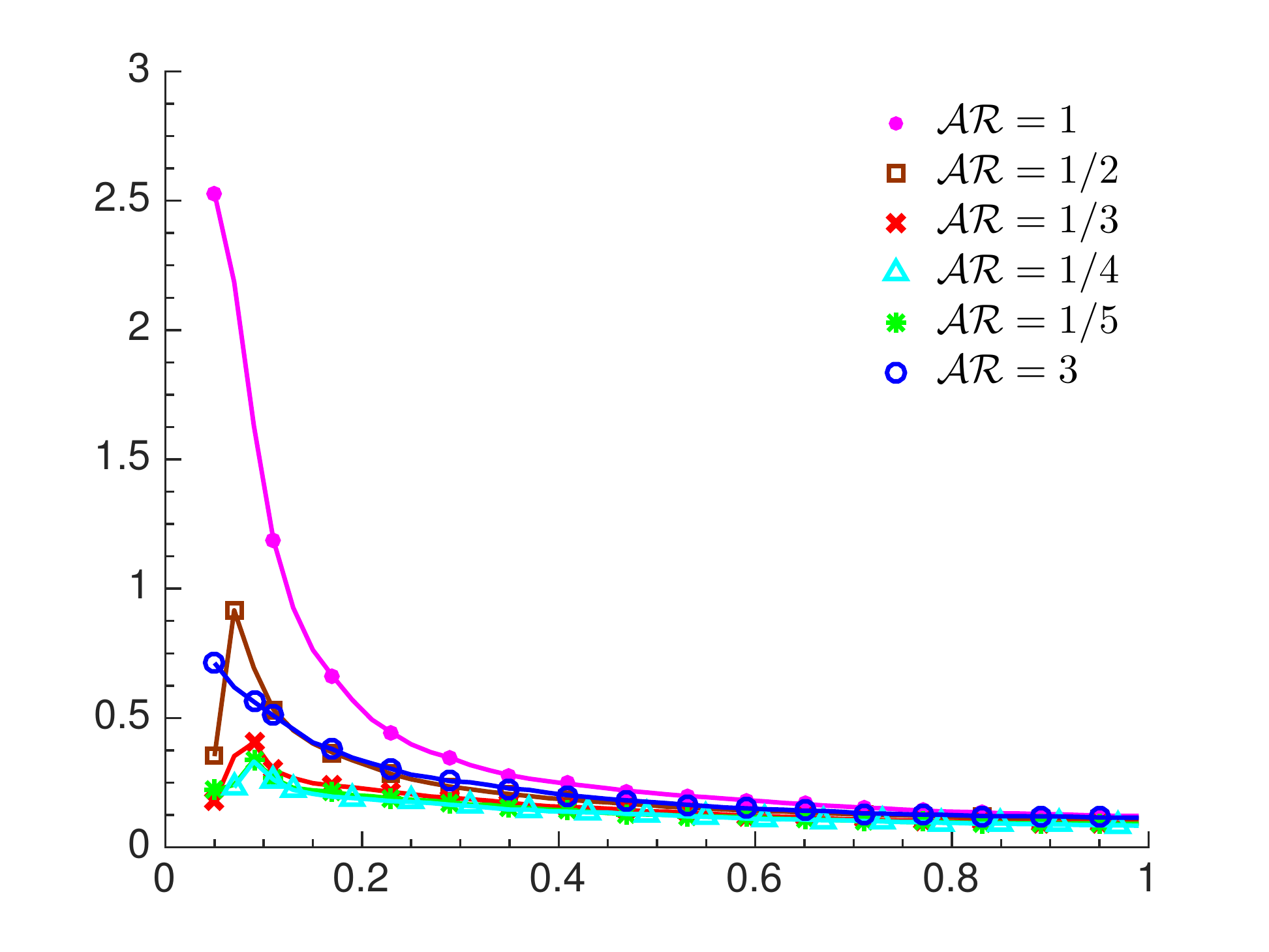}
   \includegraphics[width=0.495\textwidth]{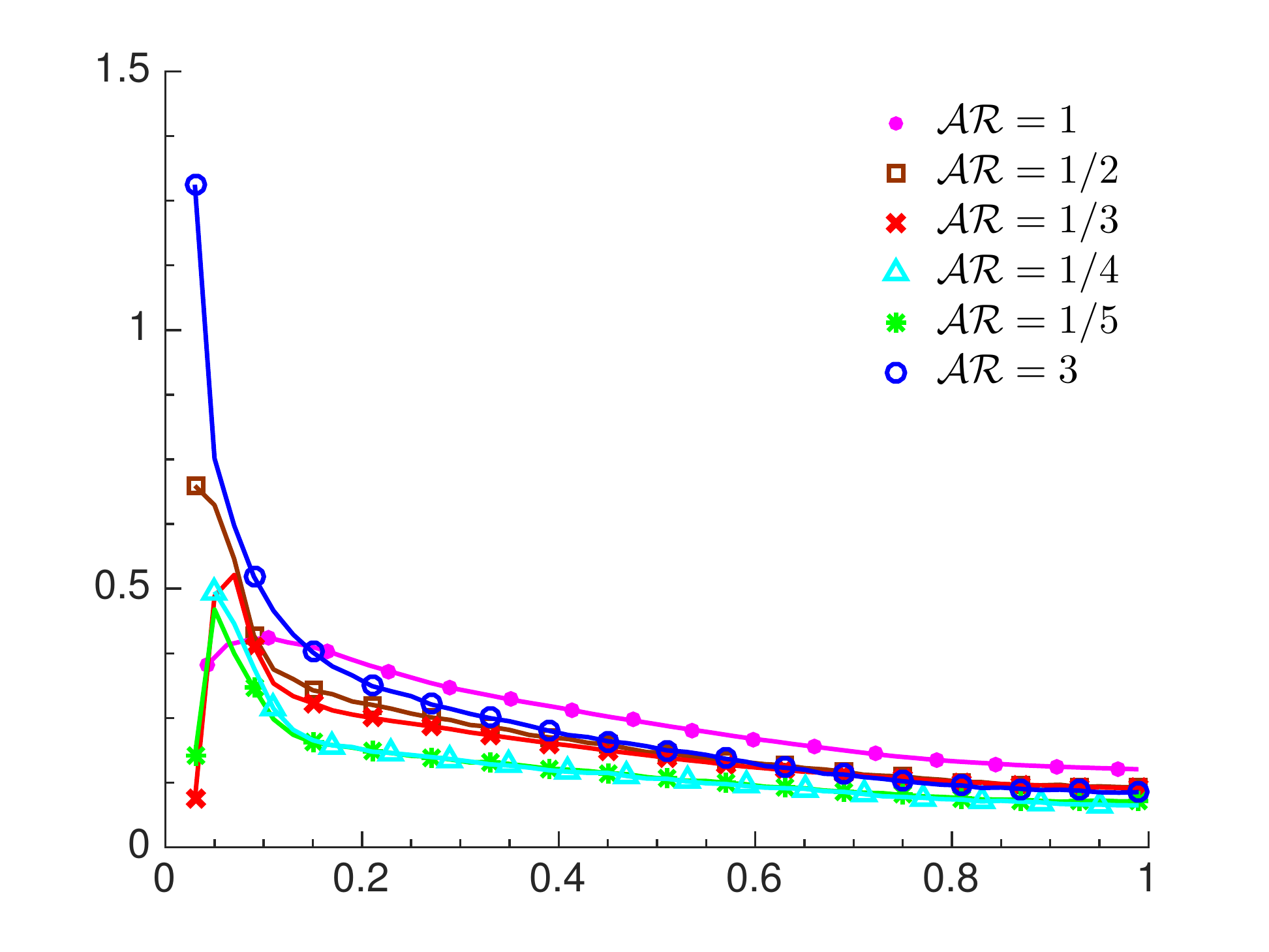}
   \put(-382,66){\rotatebox{90}{$ \Omega^z_p $}}
   \put(-193,67){\rotatebox{90}{$ {\Omega^\prime}^y_p $}}
   \put(-100,0){{$y / h$}}
   \put(-292,0){{$y / h$}}
   \put(-387,130){\footnotesize $(e)$}
   \put(-194,130){\footnotesize $(f)$} \\ 
  \caption{Profiles of solid-phase averaged data versus $y/h$; $(a)$: mean local volume fraction $\Phi$; $(b)$: cosine of the mean particle inclination angle $\theta$, measured with respect to the wall; $(c)$: magnitude of angular velocity vector $\pmb{\Omega}_p$; $(d)$: particle velocity fluctuations $\textbf{u}_p^{\prime}$; \MN{$(e)$: average of spanwise angular velocity and $(f)$: wall-normal fluctuations of angular velocity. Angular and translational velocities are normalized with $U_b/h$ and $U_b$ respectively. } }
\label{fig:Particles}
\end{figure}

Preferential sampling, particle rotation and orientation, especially close to the wall,  can lead to a strong modulation of the turbulence. Here we discuss the link between the solid-phase averaged data and the drag modifications and turbulence activity reported in the previous sections. 

The mean local volume fraction, $\Phi$, is depicted versus the wall-normal distance $y/h$ in figure~\ref{fig:Particles}$(a)$. Interestingly, a peak close to the wall is only observed for the spherical particles. This peak corresponds to a layer of particles that form due to the wall-particle interactions \citep{Picano2015} and is partially responsible for the increase of drag in this case \citep{Costa2016}. 
\MN{Non spherical-particles are observed to avoid the wall; the local volume fraction profiles for disc-like particles seem to converge for $\mathcal{AR}\leq1/3$}. To better understand this, we report in figure~\ref{fig:Particles}$(b)$ the average particle orientation throughout the channel. The orientation is measured as the cosine of the inclination angle $\theta$, which is the smallest angle between the spheroid axis of symmetry and the normal to the wall. The data clearly show the strong tendency for oblate and prolate particles to have their largest semi-axis parallel to the wall. \MN{The alignment of prolate particles with the wall in its vicinity is also observed in the studies of \cite{Eshghinejadfard2017,Wang20172}, similar to rigid fibres in the work of \cite{Do2014}}. An arbitrary symmetry axis is chosen for spheres to demonstrate the sphere random orientation. 
The magnitude of the angular velocity vector $\pmb{\Omega}_p$, \MN{normalized by $U_b/h$},  is depicted for the different aspect ratios in figure~\ref{fig:Particles}$(c)$: the angular velocities drastically decrease for oblate particles. Prolate particles, similarly to spheres, display a non-negligible angular velocity; \MN{this large value is mainly due to the wall-normal component of the angular velocity $ {\Omega^\prime}^y_p$, depicted in figure~\ref{fig:Particles}$(f)$ for different aspect ratios. Prolate particles are observed to have the largest wall-normal angular velocity in the vicinity of the wall, while this value is significantly smaller for oblate particles and decreasing with their aspect ratio. Spherical particles also experience noticeably smaller wall-normal angular velocity close to the wall, however they have the largest in the middle of the channel.} 
Conversely, in the case of spherical particles, the major contribution to the angular velocity is due to the mean spanwise component $\Omega^z_p$, as a direct consequence of the mean shear.  \MN{The mean spanwise angular velocities for different aspect ratios are given in figure~\ref{fig:Particles}$(e)$.}
Larger angular velocities result in a larger particle-induced stress as the rotation of particles larger than the smallest scale of the flow opposes stretch and deformation. To better distinguish the effect of mean flow and fluctuations on the dynamics of particles, we have simulated the behavior of single particles kept at a fixed distance from the wall ($y/h=0.1$) in the laminar flow defined by the turbulent profile.
These additional runs (not shown in the figure) show that a sphere tends to a constant angular velocity in the spanwise direction, oblate particle with $\mathcal{AR}=1/2$ exhibits a periodic spanwise angular velocity, with the majority of time spent parallel to the wall, and, more interestingly, the motion of particles with $\mathcal{AR}=1/3$, $1/4$, $1/5$ and $3$ converges to zero angular velocity with the broad side approximately parallel to the wall (slightly tilted). 
The tendency of non-spherical particles to remain parallel to the wall is therefore a mean flow effect and quenches effectively the velocity fluctuations, resulting in turbulence attenuation. This mechanism is more effective when the particles are larger than the turbulence structures. The role of the particle size can be deduced from the results in figure~\ref{fig:Particles}: local volume fraction, particle rotation and orientation attain similar values for oblates with $\mathcal{AR}\leq1/3$, however the turbulence attenuation increases for broader oblates. This can also be concluded from figure~\ref{fig:Particles}$(d)$, where the magnitude of the particle velocity fluctuations, \MN{normalized by $U_b$} is shown to decrease with $\mathcal{AR}$. \MN{The local velocity of the solid phase are computed, taking into account the rigid body motion of the particles, $\textbf{u}_p + \pmb{\Omega}_p \times (\textbf{x} - \textbf{x}_c)$, with $\textbf{x}$ and $\textbf{x}_c$ denoting an arbitrary point inside the particle and the particle center respectively.}

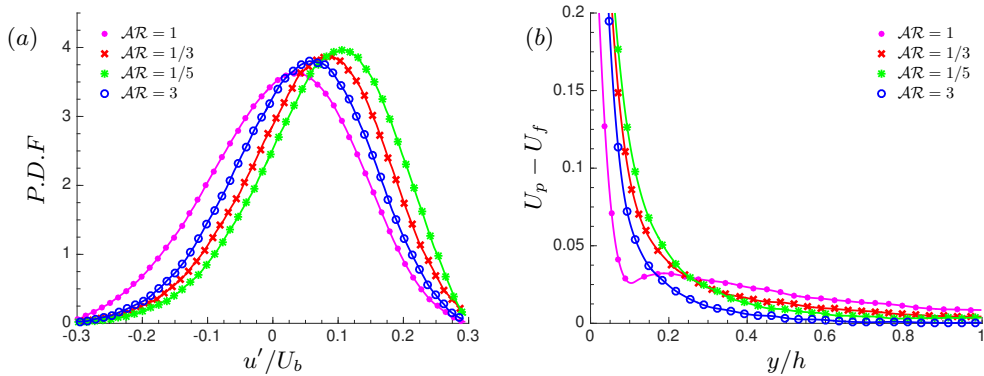
\begin{figure}
  \centering
   \includegraphics[width=0.495\textwidth]{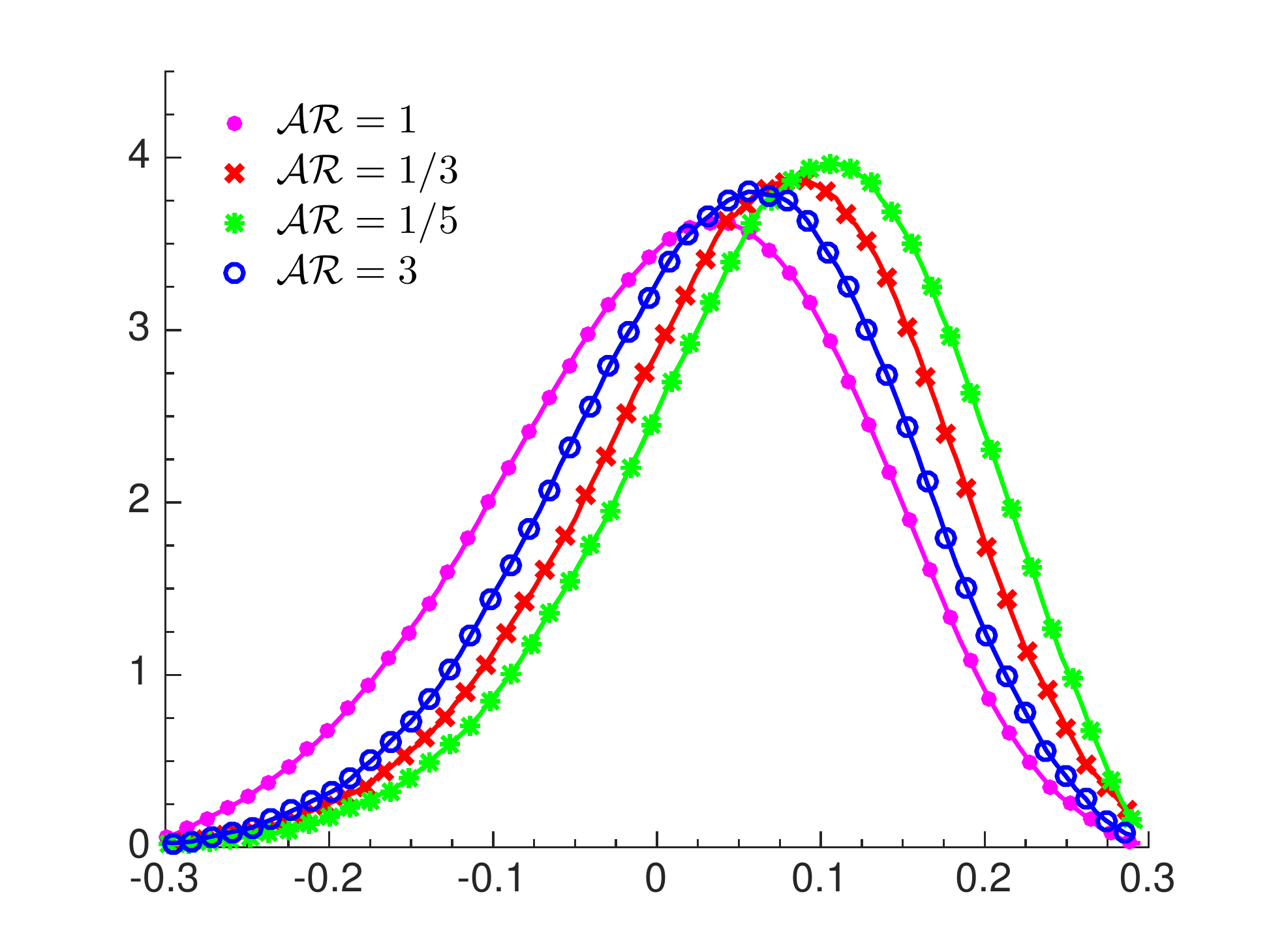}
   \includegraphics[width=0.495\textwidth]{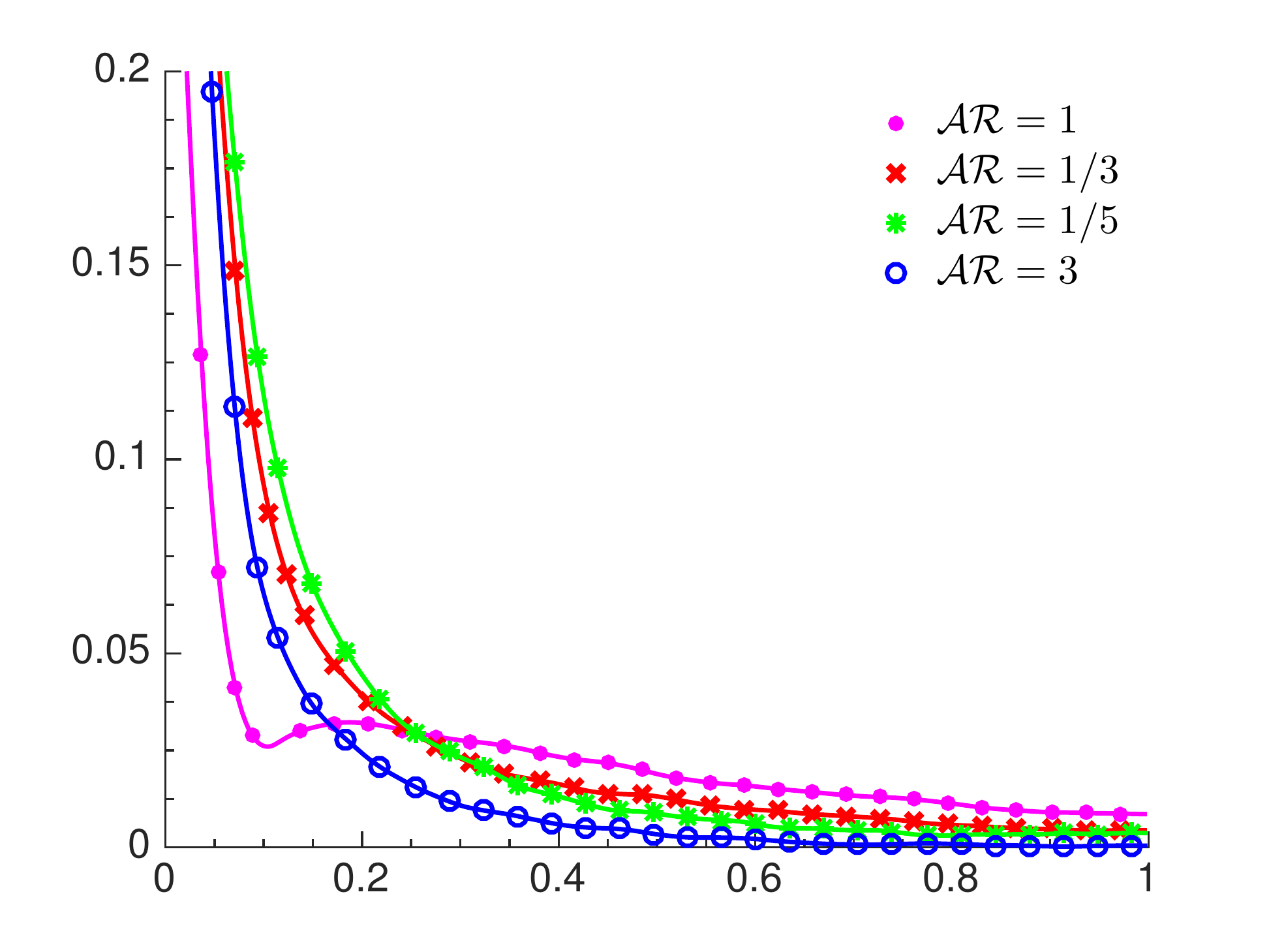}   
   \put(-377,60){\rotatebox{90}{$P.D.F$}}
   \put(-190,58){\rotatebox{90}{$U_p - U_f$}}
   \put(-99,-1){{$y / h$}}
   \put(-296,-1){{$u^{\prime} / U_b$}}
   \put(-385,120){\footnotesize $(a)$}
   \put(-190,120){\footnotesize $(b)$}
  \caption{$(a)$ Probability density function (PDF) of streamwise fluid velocity fluctuation, surrounding particles close to the wall \MN{($0.1<y/h<0.2$)} and $(b)$: the difference between the particle and the fluid mean velocity, \MN{normalized by the bulk velocity $U_b$}.}
\label{fig:pdf}
\end{figure}

Finally, we examine the probability density function (PDF) of the streamwise fluid velocity fluctuations in the vicinity of the particles and in a region close to the wall \MN{($0.1<y/h<0.2$)}, see figure~\ref{fig:pdf}$(a)$. The fluid velocity fluctuations relative to each particle are the average on the spheroidal shell $5\%$ larger than the particle itself. The data indicate a clear tendency for particles to sample high-speed regions close to the wall. 
This observed preferential sampling is weakest for spheres, followed by prolates and more and more evident when decreasing the aspect ratio of oblate particles. Figure~\ref{fig:pdf}$(b)$ shows the difference in the mean particle $U_p$ and fluid $U_f$ velocity: the difference in the region $y/h<0.2$ is consistent with the PDF reported  in figure~\ref{fig:pdf}$(a)$. \MN{It should be noted that the larger mean velocity of the particles, compared to the fluid phase does not necessarily indicate a large slip velocity as particles are shown to have a tendency to sample the high speed regions close to the wall. The larger mean velocity of the particles close to the wall is consistent with the results of \cite{Picano2015} for spherical particles and \cite{Do2014,Mehdi2017,Eshghinejadfard2017} for non-spherical ones.}

\section{Final remarks}

We have reported results from the simulations of turbulent channel flow of suspensions of finite-size spheroidal particles with different aspect ratios $\mathcal{AR} = 3$, $1$, $1/2$, $1/3$, $1/4$ and $1/5$ at volume fraction $\phi = 10\%$. By computing the contribution of the viscous, Reynolds and particle-induced shear stresses to the momentum transfer we show turbulence attenuation (with respect to the single-phase flow) for all the studied cases, except for spheres where an increase in the turbulence activity is observed. The turbulent Reynolds stresses reduce with the aspect ratio of oblate particles, i.e. for flatter disc-like particles. Overall drag reduction with respect to the single phase flow is observed for the oblate particles with $\mathcal{AR}\leq1/3$, once the turbulence attenuation is large enough to compensate the particle-induced stresses;  in the cases with $\mathcal{AR} = 1/2$ and $3$, on the contrary, the reduction in the turbulence activity (Reynolds shear stress) is lower than the additional stress induced by the particles. \MN{Particle-induced stress is found to be a function of particle rotation, since rotation comes as a consequence of opposing stretch and deformation, and therefore increasing the additional stress induced by the solid phase.}
In particular, the drag is shown to increase significantly for spherical particles and slightly for oblates with $\mathcal{AR} = 1/2$, while no relevant drag modification is reported for the case of prolate particles. Different mechanisms are found in this study to explain how the dynamics of spherical, oblate and prolate particles modulate the turbulence close to the wall.

Spheres form a layer close to the wall, rotating with high spanwise angular velocity. Their rotation brings high momentum flow towards the wall, increasing the strength of sweep events (figure~\ref{fig:Quad}$a$) and therefore increasing the turbulence activity. \MN{Flow filed in the vicinity of the particles is shown in figure~\ref{fig:Quad}$c$ to generate strong sweep events close to the wall.} The presence of a particle-layer close to the wall also results in noisier low and high-speed streaks  (figure~\ref{fig:Corr}$b$).  

Prolate particles are oriented with their major axis parallel to the wall, experiencing a significant wall-normal angular velocity due to the velocity fluctuations. Their rotation creates momentum mixing in the streamwise-spanwise plane, which results in the disruption of the near-wall low-speed streaks, see figure~\ref{fig:Corr}$d$. In this way,  prolate particles reduce the peak of streamwise and spanwise velocity fluctuations and reduce the wall-normal fluctuations as they resist rotations in the spanwise direction.

Oblate particles tend to be parallel to the wall, while also avoiding its vicinity and experiencing significantly smaller angular velocities with respect to the other shapes considered. They sample high-speed regions close to the wall, increasing the spacing between the low-speed streaks (figure~\ref{fig:Corr}$c$). These particles create a strong shield that decreases the turbulent activity by resisting rotation and therefore dampen the velocity fluctuations. Preferential sampling, rotation and orientation of oblate particles converge to similar values for oblates with $\mathcal{AR}\leq1/3$, however the turbulence attenuation continues to increase due to their larger major-axis.

\section*{Acknowledgements}
This work was supported by the European Research Council Grant No. ERC-2013-CoG-616186, TRITOS. The authors acknowledge computer time provided by SNIC (Swedish National Infrastructure for Computing) and the support from the COST Action MP1305: Flowing matter.
\bibliographystyle{jfm}
\bibliography{AR_Paper}

\end{document}